\documentclass[natbib]{svjour3}                     
\smartqed  
\usepackage{graphicx,graphics}
\usepackage{amssymb}
\newcommand{\te}{$T_e$\,}

\newcommand{\ti}{$T_i$\,}
\newcommand{\kms}{km\,s$^{-1}$\,}
\newcommand{\tetp}{\rm $T_e/T_p$}
\newcommand{\memp}{\rm $m_e/m_p$}

\newcommand{\vs}{\rm $V_s$}
\newcommand{\chan}{{\it Chandra}}
\newcommand{\bn}{\rm $I_b/I_n$}

\begin{document}

\title{Electron-Ion Temperature Equilibration in Collisionless Shocks: the Supernova Remnant-Solar Wind
Connection
}

\titlerunning{Equilibration in Collisionless Shocks}        

\author{Parviz Ghavamian         \and
        Steven J. Schwartz  \and
        Jeremy Mitchell \and
        Adam Masters  \and
        J. Martin Laming 
}


\institute{Parviz Ghavamian \at
              Department of Physics, Astronomy and Geosciences \\
	      Towson University\\
              Towson,Maryland, USA, 21252\\
              Tel.: +1(410)-704-3137\\
              Fax: +1 (410)-704-3700\\
              \email{pghavamian@towson.edu}           
           \and
           Steven J. Schwartz \at
              Department of Physics\\
	      Imperial College London\\
	      Tel.: +44(0)20-7594-7660\\
	      Fax: +44(0)20-7594-7772\\
	      \email{s.schwartz@imperial.ac.uk} \\
           \and
              Jeremy Mitchell \at
              Department of Physics\\
              Imperial College London\\
              Tel.: +44(0)20-7594-7660\\
              Fax: +44(0)20-7594-7772\\
              \email{j.mitchell@imperial.ac.uk} \\
\and
              Adam Masters  \at
              Institute of Space and Astronautical Science\\
              Japan Aerospace Exploration Agency\\
	      Kanagawa 252-5210, Japan\\
              Tel.: +81-50-3362-6079\\
              \email{a.masters@stp.isas.jaxa.jp} \\
\and
              J. Martin Laming  \at
              Space Science Division \\
              Naval Research Laboratory\\
              Code 7684, Washington, DC 20375\\
              Tel.: +1 (202) 767-4415 \\
              \email{j.laming@nrl.navy.mil}
}

\date{Received: date / Accepted: date}

\maketitle

\begin{abstract}
Collisionless shocks are loosely defined as shocks where the transition between pre-and post-shock
states happens on a length scale much shorter than the collisional mean free path. In the absence
of collision to enforce thermal equilibrium post-shock, electrons and ions need not have the same
temperatures. While the acceleration of electrons for injection into shock acceleration processes
to produce cosmic rays has received considerable attention, the related problem of the shock
heating of quasi-thermal electrons has been relatively neglected.

In this paper we review that state of our knowledge of electron heating in astrophysical shocks,
mainly associated with supernova remnants (SNRs), shocks in the solar wind associated with
the terrestrial and Saturnian bowshocks, and galaxy cluster shocks. The solar wind and SNR samples indicate 
that the ratio of electron temperature,
($T_e$) to ion temperature ($T_p$) declines with increasing shock speed or Alfv\'en Mach number.
We discuss the extent to which such behavior can be understood via cosmic ray-generated
waves in a shock precursor, which then subsequently damp by heating electrons.  Finally, we
speculate that a similar mechanism may be at work for both solar wind and SNR shocks.

\end{abstract}

\section{Introduction}
\label{intro}
Shock waves have been observed in a wide range of environments outside the Earth, from the solar wind to the hot gas in galaxy clusters.  However, the mechanism whereby the gas in these environments is shocked has been poorly understood.   While shock transitions in the Earth's atmosphere are mediated by molecular viscosity (and hence direct particle collisions), those in interstellar space and the solar wind are too dilute to form in this way.  In non-relativistic shocks, the role of collisions is effectively played by collective interactions of the plasma with the magnetic field.  This
results in a multi-scale shock transition having sub-structure at ion kinetic length scales 
(Larmor radius or inertial length) and potentially electron kinetic scales (inertial lengths or whistler mode) (e.g., Schwartz et al. 
(this volume), Treumann 2009).
Such plasmas are termed collisionless.  The magnetic fields threading through the charged particle plasmas in space endow the plasmas with elastic properties, much like a fluid.   The kinetic energy of the inflowing gas is dissipated within this fluid via collective interactions between the particles and magnetic field, transferring energy from the magnetic field to the particles.  
The collective processes are the result of the DC electromagnetic fields present in the shock transition layer, kinematic phase mixing, and also plasma instabilities; the last give rise to a rich range of plasma waves and turbulent interactions.

It has long been known that these processes may heat the electrons beyond the mass-proportional value predicted by the Rankine-Hugoniot jump conditions - ample evidence is found in spacecraft studies of solar wind shocks (Schwartz et al. 1988) and multi wavelength spectroscopy of supernova remnants (Ghavamian et al. 2001, 2002, 2003, 2007; Laming et al. 1996; Rakowski et al. 2008) and galaxy cluster gas (Markevitch et al. 2005; Markevitch \& Vikhlinin 2007;  Russell et al. 2012) understanding how this process depends on such shock parameters as shock speed, preshock magnetic field orientation and plasma beta has been slow.

In collisionless plasmas, the downstream state of the plasma cannot be uniquely determined from the upstream parameters because
the Rankine-Hugoniot jump conditions only predict the {\it total} pressure downstream, not the individual contributions from the electrons
and ions: $P\,=\,n_i\,k\,T_i\,+\,n_e\,k\,T_e$.   At the limit of a strong shock, $n_e$ and $n_i$ are each 4 times their preshock values, so
the relative values of \te and \ti immediately behind the shock are wholly dependent upon the nature of the collisionless heating processes
occurring at the shock transition.    Although an MHD description can be used to describe the behavior of the gas far upstream and far downstream of the shock, a more detailed kinetic approach is required for understanding how the dissipation at the shock front transfers
energy from plasma waves and turbulence to the electrons and ions.

Non-relativistic collisionless shocks can be broadly sorted into three categories: slow, intermediate and fast.  The three types are defined according to the angle between the shock velocity and upstream magnetic field, as well as the relative value of the shock speed compared to the upstream sound speed ($c_s\,\equiv\,\sqrt{\gamma\,P/\rho}$) and Alfv\'en speed ($v_A\,\equiv\,B/\sqrt{4\pi\,\rho_i}$).  Most astrophysical shocks are quasi-perpendicular (i.e., they propagate at a nearly right angle to the preshock magnetic field), allowing only for fast-mode propagation.  In that
case, the relevant quantity is the magnetosonic Mach number, $M_{ms}$ ($\equiv\,v_{sh} / \sqrt{v_A^2\,+\,c_s^2}$).   Collisionless shocks have
also been classified according to whether the flow speed exceeds the sound speed in the downstream plasma (Kennel 1985).  Above the critical Mach number where the flow is subsonic, the dissipation of flow energy into thermal energy can no longer be maintained by electrical resistivity,
and plasma wave turbulence (the cause of
which are instabilities generated when the electron and ion distribution functions are distorted at the shock transition) is required (Kennel 1985).  Shocks above
the critical Mach number are termed supercritical, while those below are termed subcritical.  Note that even for subcritical shocks, observations suggests that kinetic processes other than resistivity and turbulence contribute to the shock dissipation (Greenstadt and Mellott 1987).There are also other, higher critical Mach numbers related to the formation of subshocks (Kennel et al. 1985) and non-steady cyclic shock reformation beyond the whistler critical Mach number (Krasnoselskikh et al. 2002).  The critical Mach number for quasi-perpendicular shocks
is estimated to be rather low $\sim$2.8 (Edmiston and Kennel 1984). The Mach numbers of most SNR shocks are expected to be well in excess of this value,
meaning that they are both fast-mode and supercritical.

As one approaches the high Mach numbers expected for astrophysical shocks ($M_A\,\sim\,$20-100 for an ambient magnetic field of 3 $\mu G$), a 
greater and greater fraction of the
incoming ions become reflected back upstream from the shock front, corresponding to an increasingly turbulent and and disordered shock
transition.  Physical parameters such as $B$, $T$ and $n$ no longer jump in an ordered manner (i.e., the transitions are no longer laminar).
In addition to the reflected particles, the hot ions from downstream become hot enough to escape upstream, further enhancing the
population of ions in front of the shock.  These ions, which form a precursor, are now believed to play an essential role in the dissipation
of high Mach number collisionless shocks.  Aside from providing the seed population for the acceleration of cosmic rays, the precursor
ions are likely generate a variety of plasma waves capable of selectively heating the electrons over the ions, thereby providing an important
mechanism for raising \tetp\, above the mass-proportional value of $m_e/m_p$.

\section{Formalism: Equilibration Timescales}

Taken at face value, the Rankine-Hugoniot jump conditions predict that electrons and ions
will be heated in proportion to their masses:

\begin{equation}
k \,T_{e,i}\,=\,\frac{3}{16}{m_{e,i}}\,V^2_{sh}
\label{massprop}
\end{equation}

In collisionless shocks, there are three relevant timescales to consider: the time required
for Coulomb collisions to isotropize a distribution of electrons, $t_{ee}$,
the time required for Coulomb collisions to isotropize distribution of ions, $t_{ii}$,
and finally the time required for the electrons and ions to equilibrate to a common temperature,
$t_{ei}$ (Spitzer 1962). After a time scale $t_{ee}$ or $t_{ii}$, the particles in question
attain a Maxwellian velocity distribution. The self-collision time, $t_{c,ee}$ for electrons of
density $n_e$ and temperature $T_e$ is (Spitzer 1962):

\begin{equation}
t_{c,ee}\,=\,\frac{0.266\,T_e^{3/2}}{n_e\,ln\,\Lambda_e}\,\,sec\,
\approx\,\frac{0.0116\,V_{s}(1000)^3}{n_e\,ln\,\Lambda_e}\,\,\,yr
\label{relax}
\end{equation}
where ln$\Lambda$ is the Coulomb logarithm, Equation \ref{massprop} has been used to
write the relaxation time in terms of the shock speed and $V_s(1000)$ is the shock speed in units of 1000 \kms.
For a young SNR having $V_s\,\sim\,$1000 \kms, postshock density n$\,\sim\,$1 cm$^{-3}$ and ln $\Lambda\,\sim$30 the time required
to establish a Maxwellian distribution at the electron temperature given by equation 1 is $t_{c,ee}\,\sim\,$ 0.004 yrs.  The electrons are isotropized by self-collisions first, then the protons, and finally
over a longer timescale the electrons and protons equilibrate to a common temperature.  For this reason, Coulomb
collisions alone are unable to establish equilibration electron and proton distributions at the shock front.
This equilibration  is described by the relation

\begin{equation}
\frac{d T_e}{dt}\,=\,\frac{T_p\,-\,T_e}{t_{c,ep}}
\label{tetp}
\end{equation}
where
\begin{equation}
t_{c,ep}\,=\sqrt{m_p\over m_e}t_{c,pp}\,={m_p\over m_e}t_{c,ee}
\label{tep}
\end{equation}

The temperatures $T_e$ and $T_i$ equilibrate to a common density-weighted average temperature $T_{av}$, given by
$T_{av}\,=\,\frac{3}{16}\mu m_p V_s^2$,
where $\mu$ is the mean mass per particle ($=\,(1.4/2.3)\,=\,0.6$ for cosmic abundances).
For mass-proportional heating, the timescale given by Equation \ref{tep} is $\sim $2000 yrs for \vs\, $ \geq $1000 \kms, of similar order but longer than the proton-proton isotropization timescale, $t_{c,pp}$.
These are longer than the age of the SNR, substantially so at higher \vs\, indicating that for minimal heating (i.e., \tetp\,$=\,\frac{m_e}{m_p}\,\sim\,$1/1836)
the electrons and ions will not equilibrate to $T_{av}$ during the lifetime of the SNR (Itoh 1978; Draine \& McKee 1993).

The arguments above indicate that Coulomb collisions are ineffective at both isotropizing the heavy ion particle distributions
and equilibrating the electron and ion temperatures at the transtions of collisionless shocks.  The emission spectra of non-radiative SNRs
(dominated mostly by X-ray and ultraviolet emission) should therefore be sensitive probes of the collisionless heating processes
at the shock transition.  We consider the observational constraints of these processes below.

\section{Observational Constraints From SNRs}
\label{sec:1}

The most useful shocks for studying collisionless equilibration processes are those exhibiting detectable emission
from the immediate postshock gas.  To be diagnostically useful, the emission should arise from the region close to
the shock front, where temperature disequilibrium between electrons, protons and heavy ions is substantial enough to affect both
the relatives fluxes and relative velocity widths of emission lines.  The shape and extent of the spatial profile
for the different emission species behind the shock is also a useful diagnostic, especially for the UV resonance
lines of He~II $\lambda$1640,  C~IV $\lambda\lambda$1548, 1550, N~V $\lambda\lambda$1238,1243, and O~VI $\lambda$1032, 1038.
For a given shock speed, the distance behind the shock where the emission peaks depends strongly on the initial electron
temperature, and  electron temperature immediately behind the shock.

The ubiquity of non-radiative SNRs, as well as their relatively simple geometry and very high shock speeds,
makes these objects the most important laboratories for investigating the efficiency and nature of electron-ion
and ion-ion equilibration.  Other non-radiative shocks available for study are those occurring in stellar wind
bubbles (for example, Wolf-Rayet bubbles) (Gosset et al. 2011) and in galaxy cluster shocks (e.g., the 'Bullet Cluster' (Markevitch
et al. 2005) and Abell 2146 (Russell et al. 2012)).  However, these shocks are far less frequently observed than those in SNRs,
leaving the latter as the most important objects providing both a broad range of shock speeds and corresponding diagnostic information (such as
proper motions and spatially resolved structure).  Shocks in star forming regions, such as HH objects and their associated
bow shocks, tend to be clumpy and complicated, and are usually located in dense environments ($n\,\sim\,$100-1000 cm$^{-3}$)
with low enough shock speeds (\vs\,$\sim\,$100-200 \kms) to be radiative.  In those cases, the optical and UV emission (where the
most valuable shock diagnostic line emission arises) is dominated by emission from the cooling and recombination zones far
downstream from the shock.   In those regions the cooling of the gas below 100,000K and the accompanying compression of the
gas result in a collisional plasma with $T_e\,=\,T_p\,=\,T_i$.  This erases any 'memory' of the initial electron-ion
equilibration.  Furthermore, a significant fraction of the Ly $\alpha$ continuum produced in the recomination zone is
expected to pass upstream and ionize the preshock gas.  In strong shocks (\vs\,$>$80 \kms), this results in
complete ionization of hydrogen (Shull \& McKee 1979; Cox \& Raymond 1985; Dopita et al. 1993), thus precluding the
use of collisionally excited Balmer line emission from neutral H (described in the next section) as a temperature
equilibration diagnostic.

\subsection{Optical Spectroscopy of Balmer-Dominated Shocks: The \tetp\,$\propto$\,V$_{sh}^{-2}$ Relation}
\label{sec:2}

\begin{figure*}
  \begin{center}
  \includegraphics[scale=0.6]{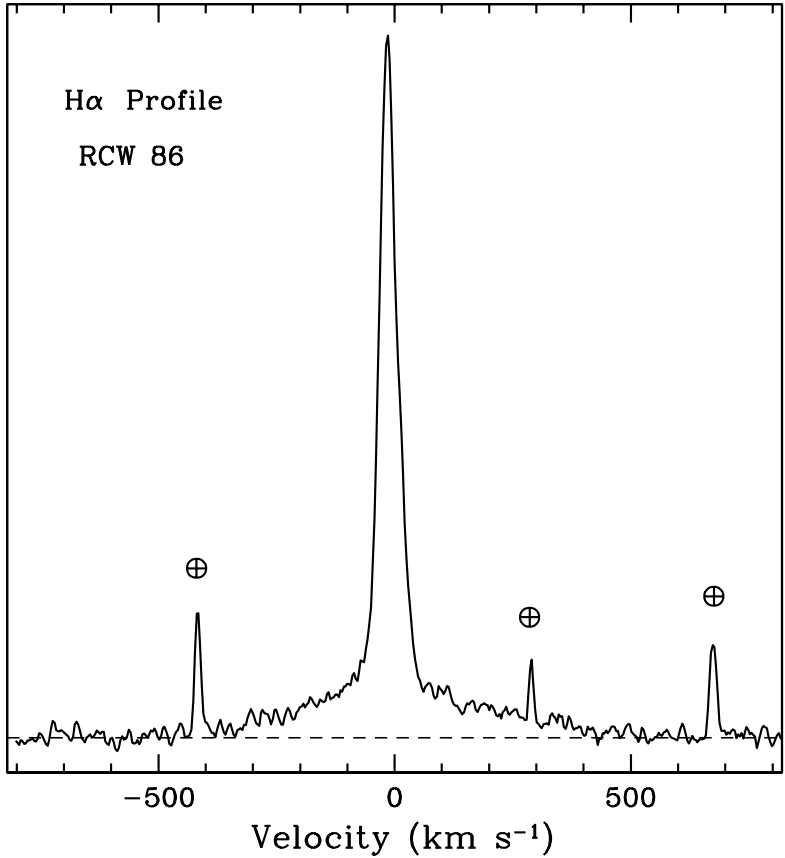}
  \end{center}
\vskip -0.2in
\caption{The H$\alpha$ line profile of a Balmer-dominated shock in RCW 86, acquired with
a high dispersion (resolution $\sim$ 10 \kms) spectrograph (Sollerman et al. 2003).  The broad
($\sim$ 500 \kms) and narrow ($\sim$30 \kms) H$\alpha$ lines are both resolved.  Night sky
lines (marked by the circles) have been left in to demonstrate their noticeably narrower width
compared to the H$\alpha$ lines}
\label{fig:balmerprofile}       
\end{figure*}

In the late 1970s it was discovered that very fast ($\sim$2000 \kms) shocks in young SNRs
could generate detectable optical emission very close to the shock transition (Chevalier \& Raymond 1978;
Chevalier, Kirshner \& Raymond 1980), providing a valuable diagnostic of physical conditions at
the shock before Coulomb collisions or cooling could alter them.  This emission is produced by collisional
excitation of H~I as it flows into the shock front.  The cold neutral component does not interact directly with the plasma waves
and turbulence at the shock, while the ionized component is strongly heated and compressed by a factor of four (when the
shock is strong).  Some of the cold H is destroyed by collisional ionization; however, the rest of the cold H undergoes
charge exchange with hot ions behind the shock, generating a separate population of hot H.  Approximately 1 in every 5
collisions results in collisional excitation to the n=3 level of H, producing H$\alpha$ and Ly $\beta$ emission.  The H$\alpha$
line from the cold neutrals is narrow and reflects the preshock temperature ($\leq$30,000~K), while that from the hot neutrals
is broad (typically $\geq$500 \kms), and reflects the postchock temperature (and to a large extent, the velocity distribution) of the protons
(Chevalier \& Raymond 1978; Chevalier, Kirshner \& Raymond 1980; Smith et al. 1991; Ghavamian et al. 2001) (Figure \ref{fig:balmerprofile}).
In Balmer-dominated shocks,the broad to narrow H$\alpha$ flux ratio is proportional to the ratio of the charge exchange rate to the
ionization rate, with the latter being highly sensitive to the electron and proton temperatures.
This makes the broad to narrow ratio, \bn, very sensitive to \tetp.  There is only weak dependence of \bn\, on the preshock H~I fraction and preshock
temperature, mainly due to differences in the amount of Ly $\beta$ converted into H$\alpha$ in the narrow component (Ghavamian et al. 2001;
van Adelsberg et al. 2008).

The first systematic attempt to use Balmer-dominated SNRs to infer \tetp\, for collisionless shocks was attempted by
Ghavamian et al. (2001).  Using the broad H$\alpha$ line widths measured from a sample of Balmer-dominated shocks, they
constrained the range of plausible shock speeds between the limits of minimal equilibration and full equilibration.  They then
predicted the broad-to-narrow ratios for a grid of shock models over this range of \vs\, and \tetp, allowing \tetp\, to be constrained.
The range of broad component H$\alpha$ widths observed in SNRs ranges from $\sim$250 \kms\, for the slowest Balmer-dominated shocks (Cgynus Loop),
to $\sim$ 500 \kms\, for intermediate-velocity shocks (RCW 86) and finally $\sim$2600 \kms\, for the fastest shocks (SNR 0509$-$67.5).
This corresponds to a well-sampled range of shock speeds: nearly a factor of 10.
The primary uncertainty in measurement of the broad component width at low shock speeds ($\lesssim$200 \kms, as seen in the Northeastern
Cygnus Loop; Hester et al. 1994) is disentangling the broad and narrow components when they are of comparable width.  At high
shock speeds ($\gtrsim$2000 \kms) the main difficulty is the baseline uncertainty of the surrounding continuum: if the peak of
the broad line is low and the width very large, errors in ascertaining where the broad line merges into the background can lead
to underestimates of the broad component width.
The range of \bn\, for this sample of Balmer-dominated shocks lies between 0.4 and 1.2 (Kirshner, Winkler \& Chevalier 1987; Smith et al.
1991; Ghavamian et al. 2001; 2003, Rakowski, Ghavamian \& Laming 2009).  However,
it does not not vary with broad component width in a monotonic fashion.

Recently there has
been substantial improvement in the modeling of Balmer-dominated shocks.  Earlier calculations of the broad H$\alpha$ line profile
assumed that it formed from a single charge exchange (Chevalier, Kirshner \& Raymond 1980; Smith et al. 1991; Ghavamian et al. 2001),
and it treated the hot and cold neutrals as two separate, distinct populations, with a given neutral belonging to either one or the other.
However, in reality an interaction 'tree' is required to track the number of
photons emitted by each neutral over multiple excitations and charge exchanges.  These effects were first incorporated in the Balmer-dominated
shock models of Heng \& McCray (2007), who also found that charge exchange results in a third population of neutrals having velocity
widths intermediate between the hot and cold neutrals.
Further improvements in modeling of Balmer-dominated shocks were included by van Adelsberg et al. (2008), who included the momentum
transferred by charge exchange between the hot neutrals and protons.   This allowed the bulk velocity of the postshock neutrals to be
calculated separately from those of the protons.  Inclusion of the momentum transfer showed that for shock speeds $\lesssim$1000 \kms, charge
exchange effectively couples the fast neutral and thermal proton distributions, while for high shock speeds ($\gtrsim$5000 \kms), it
does so far less effectively.  This results in a fast neutral distribution that is skewed relative to the protons in velocity space and
an average velocity that is much higher for the fast neutrals than the protons (van Adelsberg et al. 2008).   Together, the inclusion of
all these effects has enhanced the
ability of the models to match the observed broad-to-narrow ratios (and hence predict \tetp).
In particular, the newer models can now match the low broad-to-narrow ratio (\bn\,$\approx$0.67) observed in Knot g of Tycho's SNR,
yielding \tetp\,$\approx$0.05, \vs\,$\approx$1600 \kms.  However, even after all the additional physics is included, the
Balmer-dominated shock models are still unable to reach the low broad-to-narrow ratios measured along the rims of DEM L71
(\bn\,$\approx$\,0.2-0.7; Ghavamian
et al. 2003; Rakowski, Ghavamian \& Laming 2009).  The electron-ion equilibration in DEM L 71 was instead determined via comparison of
broad H$\alpha$ line FWHM with postshock electron temperatures measured from \chan\, observations (Rakowski, Ghavamian \& 
Hughes 2003).  The most promising explanation advanced for the \bn\, discrepancy has been added narrow
component flux from the shock precursor (Raymond et al. 2011; Morlino et al. 2012), hitherto not included in the earlier shock
models (Ghavamian et al. 2003, Rakowski, Ghavamian \& Laming 2009).  These developments are described in more detail in the next
section.

The plot of \tetp\, versus \vs for the available sample of Balmer-dominated shocks shows a declining trend of equilibration with
shock speed (Ghavamian et al. 2007; Heng et al. 2007; van Adelsberg et al. 2008).   The trend is described by Ghavamian et al. (2007)
as full equilibration for shock speeds up to and including 400 \kms, and a declining equilibration proportional to the inverse square
of the shock speed above 400 \kms.  This description can be characterized in the following way:
\begin{equation}
\frac{T_e}{T_p}\,=\,\left\lbrace\begin{array}{c c}{1} & {\rm if\,V_s\,< 400\,km\,s^{-1}} \\ { \frac{m_e}{m_p}\,+\,\left(1\,-\,\frac{m_e}{m_p}\right)\left(\frac{V_s}{400}\right)^{-2}} & {\rm\,\, if\, V_s\,\geq\,400\,km\,s^{-1} }\end{array}\right\rbrace
\label{tetpeq}
\end{equation}
where the functional form of the the \tetp\, relation is designed to asymptotically transition to \tetp\,=\,\memp\, at very high
shock velocities.

\begin{figure*}
  \begin{center}
  \includegraphics[width=0.55\textwidth,angle=90]{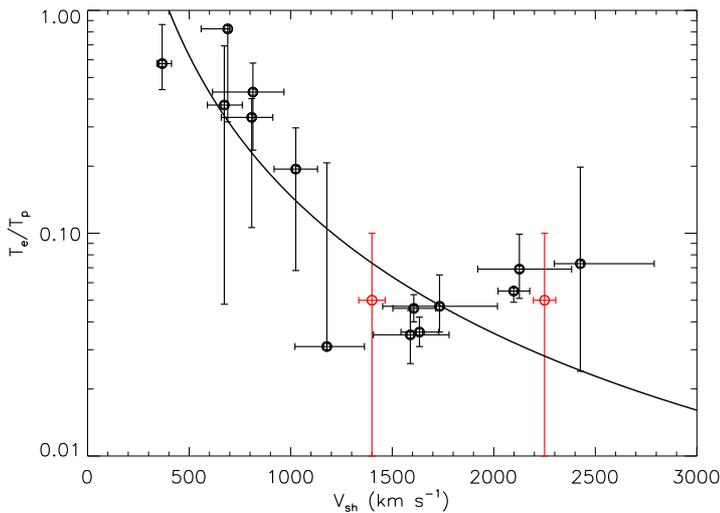}
  \end{center}
\caption{The correlation between $\beta$ ($\equiv\,T_e/T_p$) and shock speed for Balmer-dominated
SNR shocks is shown, using results from van Adelsberg et al. (2008).  The solid curve shows the $V_{S}^{-2}$
dependence inferred by Ghavamian et al. (2007) for Balmer-dominated SNR shocks. New results quoted from the NE and NW
portions of Tycho's SNR (this paper) are marked in red. The apparent upturn at \vs\,$\gtrsim$2000 \kms\, and
its associated error bars are discussed more fully in the text. }
\label{fig:plotequil}       
\end{figure*}

The most up to date plot of \tetp\, versus shock speed, reproduced from van Adelsberg et al. (2008), is shown in Figure~\ref{fig:plotequil}
(Note that the shock models used in producing these plots do not include contribution from the shock precursor).
Although the inverse correlation between \tetp\, and \vs was largely confirmed by van Adelsberg et al (2008), there may be some evidence of
departure from the \tetp\,$\propto$ $V_s^{-2}$ relation at shock speeds exceeding 2000 \kms.  van Adelsberg et al. find that when all three
measured broad component widths and broad-to-narrow ratios from SN 1006 are included in the plot (\vs\,$\sim$2200-2500 \kms), a slight
upturn in the \tetp\,-\,\vs\, relation appears.  The \tetp\, ratios for those cases are found to be $\sim$0.03, superficially similar
to {\rm $\sqrt{m_e/m_p}$}, rather than ${\rm m_e/m_p}$.  The reason for the discrepancy is not clear, nor whether this indicates a
breakdown in the $V_s^{-2}$ dependence at shock speeds exceeding 2000 \kms.   Some caveats to consider when interpreting the
upturn in \tetp\, seen in Figure \ref{fig:plotequil} are as follows: For shock speeds $\gtrsim$2000 \kms\,
collisional ionization and excitation of H are primarily caused by proton (and to a lesser degree, alpha particle) impact (Laming et al. 1996;
Ghavamian 1999; Ghavamian et al. 2001; Tseliakhovich et al. 2012).  Experimentally measured cross sections for these interactions have still
not been available to high precision (uncertainties $\sim$20\%-30\% still exist), although more sophisticated theoretical calculations
are now becoming available (see, for example, Tseliakhovich et al. 2012) .  In addition, as the broad component width increases, the H$\alpha$
profiles are spread out over an increasing number of pixels, resulting in noisier spectra and greater measurement uncertainty in the broad
component width.  These larger error bars in turn result in a larger uncertainty in \vs, especially at shock speeds of 2000 \kms\, and higher.

\begin{figure*}
  \begin{center}
  \includegraphics[width=0.75\textwidth]{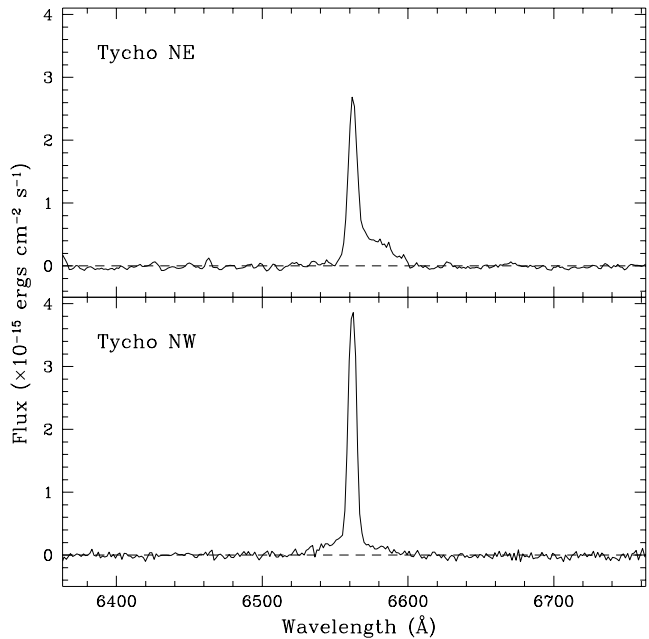}
  \end{center}
\caption{H$\alpha$ profiles from two locations along the Balmer-dominated rim of Tycho's SNR (Ghavamian 1999).
Top: the spectrum of a clump immediately to the north of Knot g (\bn\,=\,0.85$\pm$0.04; FWHM = 1300$\pm$65 \kms).
The broad component is substantially redshifted from the center of the narrow component (+10.7$\pm$1.2 \AA,
or 490$\pm$55 \kms), indicating that the clump is located on the far side of the shell, projected $\sim$20$^{\circ}$ inside
of the main Balmer-dominated rim.  Bottom: profile acquired from the northernmost filament in Tycho, \bn\,=\,0.45$\pm$0.15,
FWHM = 2040$\pm$55 \kms.
   }
\label{fig:tychocennwspec}       
\end{figure*}

The best way to further constrain the equilibration-shock speed relation is to add new data points to the curve shown in
Figure~\ref{fig:plotequil}.  To this end, we present Balmer-dominated H$\alpha$ profiles for two additional positions in Tycho's
SNR (different from those of Knot g presented by Kirshner, Winkler \& Chevalier (1987), Smith et al. (1991) and Ghavamian
et al. (2001)).  These profiles, shown in Figure~\ref{fig:tychocennwspec}, were acquired with a moderate resolution spectrograph in 1998
(for details on the observational setup, see Ghavamian 1999 and Ghavamian et al. 2001). The profile marked 'NE' was obtained from a clump of
H$\alpha$ emission located along the northeastern edge of Tycho's SNR, approximately 1$^{\prime}$ northward of Knot g.
The clump appears behind the main body of the Balmer filaments and exhibits a broad component that is substantially Doppler
shifted to the red (10.7$\pm$1.2 \AA, or about 490 \kms).   The Doppler shift reflects the bulk velocity of the hot posthock
proton distribution, so the significant velocity shift of the broad component centroid indicates that the shock in the NE
has a substantial velocity component into the plane of the sky, i.e., that the NE shock is located
on the far side of the blast wave shell.

The broad H$\alpha$ width of the NE shock is 1300$\pm$65 \kms\, (the smallest broad
component width measured in Tycho's SNR so far), with \bn\,=\,0.85$\pm$0.04.  The NW shock, on the other hand, has a broad
H$\alpha$ width of 2040$\pm$55 \kms\, (the largest broad component width measured in this SNR so far), with \bn\,=\,0.45$\pm$0.15.
Neither of these broad-to-narrow ratios is strictly reproduced by the latest models of van Adelsberg et al. (2008), with the
lowest predicted ratios being for Case B (the assumption of optically thick conditions for Ly $\beta$ photons in the
narrow component) and for low equilibrations (\tetp\,$\lesssim$0.1).  For these low equilibrations, the corresponding shock
speeds for the NE and NW shocks in Tycho are approximately 1400 \kms\, and 2250 \kms\, (using Figures 5 and 10 of van Adelsberg
et al. 2008).
In Figure~\ref{fig:plotequil} we have added data from the NE and NW shocks in Tycho's SNR to the \tetp\,-\vs\, plot.  The two data
points help fill in a portion of the plot where the data are sparse: the region between approximately 1200 \kms\, and 1500 \kms, as
well as the region beyond 2000 \kms, where the existing data are taken entirely from SN 1006.  The added point near 1400 \kms\,
is fully consistent with the $V_s^{-2}$ relation, while \tetp\, is not well constrained enough for the point at 2250 \kms\, to
contradict the appearance of an upturn at the highest shock velocities.  It is clear that proper characterization of the equilibration-shock
velocity curve above 2000 \kms\, will require both higher signal-to-noise spectra on existing Balmer-dominated shocks, as well
as new data points beyond a broad component width of 2500 \kms.

\subsection{Exceptions to \tetp\,$\propto$\,V$_{sh}^{-2}$}

Although the inverse squared relation between equilibration and shock velocity appears has been the most salient result of the
study of Balmer-dominated shocks, there have been discrepant results reported in a small subset of cases.
Recently the broad H$\alpha$ component in the LMC SNR 0509$-$67.5 was detected for the first time (Helder, Kosenko \& Vink 2010).
Broad emission was observed along both the northeastern rim (FWHM 3900$\pm$800 \kms) and southwestern rim (FWHM 2680$\pm$70 \kms), with the former being the
fastest Balmer-dominated shock detected to date having the characteristic broad and narrow component H$\alpha$ emission.  Interestingly, the broad-to-narrow
ratios for both shocks are exceptionally low, \bn\,=0.08$\pm$0.02 and 0.29$\pm$0.01 along the NE and SW rims, respectively.   As noted by Helder,
Kosenko \& Vink (2010) these ratios are nearly twice as large as the smallest ones predicted by the models of van Adelsberg et al. (2008),
precluding measurement of \tetp\, from the Balmer-dominated spectra and implicating excess narrow component H$\alpha$ emission in a cosmic ray
precursor once again.    Using RGS spectra of SNR 0509$-$67.5  acquired with {\it XMM-Newton}, they obtained a forward shock speed of
approximately 5000 \kms\, in the SW.   This implies a broad H$\alpha$ width of 3600 \kms, substantially smaller than the measured width of
2680 \kms.  Helder, Kosenko \& Vink (2010) suggest that this indicates  some thermal energy loss ($\sim$20\%) to cosmic ray acceleration.  This
picture is supported by the presence of nonthermal X-ray emission in their fitted RGS spectra of SNR 0509$-$67.5.

The result described above relies upon an accurate disentangling of the bulk Doppler broadening from the thermal Doppler broadening in the X-ray lines.
The disentangling depends on parameters such as the ratio of reverse shock to forward shock velocities, as well as the ratio of the gradients in
the reverse and forward shock velocities, which in turn had to be assumed from evolutionary SNR models.  The result, while intriguing, is still
significantly uncertain.  On the other hand, Helder, Kosenko \& Vink (2010) found that the X-ray shock velocities from the NE could 
only be reconciled with the observed broad H$\alpha$ width
there if \tetp\,$\approx\,$0.2, which clearly which predicts \tetp\,$\sim\,m_e/m_p$ predicted from Equation~\ref{tetpeq}.   If this result 
were to be confimed by future observations, it would present a new challenge in understanding how electron-ion equilibration occurs in fast 
collisionless shocks.  One possibility, given the presence of nonthermal X-ray emission in the spectra of 0509$-$67.5 (Warren \& Hughes 2004;
Helder, Kosenko \& Vink 2010) is that the moderate loss of thermal energy to cosmic ray acceleration may have slightly increased the compression
and reduced the temperature at the shock front compared to the case with no acceleration (Decourchelle, Ellison \& Ballet 2000; Ellison
et al. 2007).  Both of these effects would tend to render the
plasma more collisional, possibly explaining the \tetp\,$\approx\,$0.2 result.  
However, it is also worth noting that the shock velocity used for obtaining
\tetp\, in the NE is subject to the same model dependence and uncertainties as the SW measurements, so similar caution is required in its
interpretation.

A similar combined optical and X-ray study of RCW 86 was performed by Helder et al. (2009,2011).  There, the broad component H$\alpha$
widths were supplemented with electron temperatures measured from {\it XMM-Newton} RGS spectra from the same projected locations along the rim.
One of the main results of this study was that the slower shocks (broad H$\alpha$ FWHM $\sim$500-600 \kms) showed showed \tetp\,$\sim$1, agreeing with
earlier results from similar shocks observed both in RCW 86 and elsewhere (Ghavamian et al. (1999, 2001, 2007)).  However, Helder
et al. (2011) found while the results
were indeed consistent with low equilibration at the shock front for fast shocks (\tetp\,$\approx$\,0.02 for broad FWHM $\sim$1100 \kms))
and higher equilibration for the slower shocks (\tetp\,$\approx$1 for broad FWHM $\sim$650 \kms),
their X-ray derived electron temperatures were inconsistent
with $T_e\,=\,$0.3 keV at the shock front contradicting the suggestion of Ghavamian
et al. (2007) that shocks above 400 \kms\, may all heat electrons to roughly 0.3 keV.  However,
a major caveat of these results is that the forward shock in RCW 86 is believed to be impacting the walls of a wind-blown
bubble (Williams et al. 2011), resulting in substantial localized variations in shock speed around the rim.  These variations occur as different parts of the
forward shock impact the cavity wall at different times.  While the broad component H$\alpha$ widths closely trace the current
position of the shock front, the X-ray emission behind that shock arises over a much more extended spatial scale, and is sensitive
to the history of the forward shock interaction with the cavity wall.   Furthermore, narrowband H$\alpha$ imagery
of RCW 86 with the ESO Very Large Telescope (Helder et al. (2009, 2011)) shows a complex morphology of filaments, especially
along the eastern side of the SNR.  The broad H$\alpha$ components of these filaments exhibit substantial, localized variations 
in line width, ranging from $\sim$600 \kms\, FWHM
to 1100 \kms\, Ghavamian (1999), Helder et al. (2009, 2011)).  These variations reflect localized changes in density and viewing geometry
along the line of sight.  As such, uniquely mapping the observed Balmer filaments to their corresponding X-ray emission
in {\it XMM-Newton} data (especially given the somewhat coarse 10$^{\prime\prime}$ spatial resolution of that instrument) is 
fraught with uncertainty. Additional corroboration for these results would be desirable.

\subsection{Imprint of the Shock Precursor on the H$\alpha$ Line Profile}

Perhaps the most important and exciting recent development in our understanding of Balmer-dominated shocks has come
with the development of new kinetic-based shock models.  Blasi et al. (2012) have introduced a kinetic model for
following the momentum and energy
exchange between neutrals and ions, along with the back-reaction of those neutrals when they pass back upstream and form a fast
neutral precursor.  Rather than assume a Maxwellian velocity distribution for the neutrals (as had been done in previous models,
despite the lack of thermal contact between neutrals needed to justify such an assumption), both the ion and neutral distributions
are computed from their appropriate Boltzmann equations.  Building on these models, Morlino et al. (2012a,b) have confirmed what
had been suspected earlier (Smith et al. 1994;
Hester et al. 1994; Sollerman et al. 2003), namely that the broadening of the narrow component beyond the expected ISM value
($\sim$25-30 \kms\, instead of 10 \kms) is most likey due to heating in a cosmic ray precursor.  In particular, Morlino et al. (2012)
found that the characteristic charge exchange length of the incoming neutrals exceeds that of the neutrals crossing
back upstream, so that the narrow component width is impacted not by the neutral return flux, but rather by heating in the cosmic 
ray precursor.  Raymond et al. (2011)
predicted a similar broadening of the narrow component, though they focused mainly on the contribution of collisional excitation 
in the precursor to the flux in the narrow H$\alpha$ component.  Spatially resolved line broadening of the narrow H$\alpha$
component was detected in ground-based longslit spectroscopy of Knot g in Tycho (Lee et al. 2007).  In addition,
a small ramp-up in H$\alpha$ emission was observed ahead of Knot g in HST imagery Tycho's SNR (Lee et al. 2010). The
results, taken together, are strong evidence for the presence of cosmic ray precursors in Balmer-dominated SNRs.

\begin{figure*}
  \begin{center}
  \includegraphics[width=0.5\textwidth,angle=270]{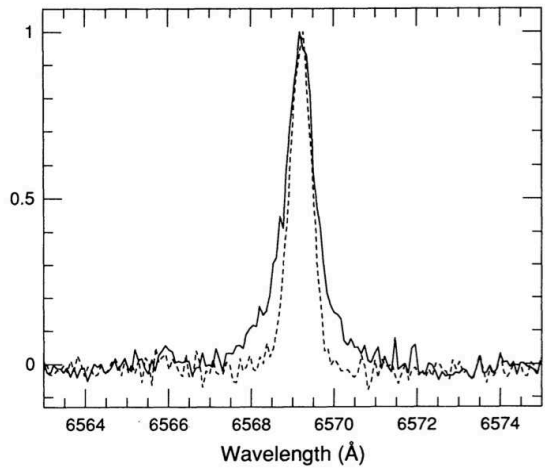}
  \end{center}
\caption{High dispersion H$\alpha$ profiles of the Balmer-dominated SNR 0509$-$67.5 in the Large Magellanic Cloud
(Smith et al. 1994). The broad component is not detected due to the high dispersion.  The narrow component
from the east rim of the SNR (solid line) shows intermediate velocity wings of width $\sim$ 100 \kms, while that
of the center (dotted line) does not.  The intermediate component is now believed to arise within a fast neutral
precursor (Morlino et al. 2012a).
   }
\label{fig:0509_3rdHalphacomp}       
\end{figure*}

One prediction of the new Balmer-dominated shock models is the existence of a third component of the H$\alpha$ emission (Morlino et
al. 2012a,b).  When the hot neutrals escape upstream, they undergo charge exchange with the colder preshock
protons.  This results in fast protons with cold neutrals, with the former rapidly equilibrating with
preshock protons and pre-heating the gas.  The temperature of the equilibrated protons in the
precursor lies between the temperature of the far upstream protons ($\sim$5000 K) and the far downstream
protons ($\sim$ 10$^6$ - 10$^8$ K), typically $\sim$10$^5$~K.  Further charge exchange between these warm protons and the preshock
neutrals gives rise to a third, 'warm' neutral component (neither fast nor slow) having velocity widths of
hundreds of \kms\,(Morlino et al. 2012a).  Interestingly, the presence of a third H$\alpha$ component was first
observationally reported by Smith et al. (1994) in their high resolution echelle spectroscopy of Balmer-dominated
SNRs in the Large Magellanic Cloud (an example of one of the spectra from Smith et al. (1994) is reproduced in 
Figure \ref{fig:0509_3rdHalphacomp}).  A third H$\alpha$ component was also reported in high resolution spectra of Knot g in   
Tycho's SNR (Ghavamian et al. 2000).
In their models of Balmer-dominated shock emission, Morlino et al. (2012a) found that the
importance of the third component relative to that of the broad and narrow components depends
strongly on the preshock neutral fraction and \tetp, in line with earlier theoretical predictions on
properties of a fast neutral precursor (Smith et al. 1994).
The fact that the third component has
been detected in Tycho's SNR (width $\sim$150 \kms) is consistent with the high preshock neutral fraction ($f_{H~I}\,\sim\,$0.9)
inferred from the broad-to-narrow ratio of Knot g by Ghavamian et al. (2001). A similar third component may have
been detected in high resolution spectra of SNR 0509$-$67.5, where measurement of the narrow component width
required the inclusion of an additiona component of width 75 \kms\, (Smith et al. 1994).
The new fast neutral precursor models predict that a substantial fraction of the H$\alpha$ excitation in Balmer-dominated shocks can
arise ahead of the shock, where warm neutrals are excited by electron impact.  Interestingly, the relative contribution of
the preshock H$\alpha$ to the total (upstream + downstream) is sensitive to \tetp\, behind the shock.  Morlino et al. (2012a) found that up to 40\% of the total H$\alpha$ flux from a Balmer-dominated shock can arise from the fast neutral precursor when
\vs\,$\sim$\,2500 \kms\, and \tetp\,=\,1 both upstream and downstream.  In these models the preshock contribution to the
total flux drops substantially for lower downstream equilibrations for \vs\,$\gtrsim$1000 \kms\, (the slowest
shocks considered by Morlino et al. 2012a).  This is generally consistent with the fact that in most cases, shock models not including
the precursor H$\alpha$ emission have been able to match the observed broad-to-narrow ratios.  In other words, if the postshock
temperature equilibration were not low for such fast shocks, the agreement between the observed and predicted \bn would have been
substantially worse for such remnants as Tycho's SNR and SN 1006.

\subsection{Ultraviolet and X-ray Studies of Balmer-Dominated Shocks}
\label{sec:3}

SN 1006 is an example of a SNR accessible to UV spectroscopy due to its galactic location, 450 pc
up from the galactic plane, and therefore with relatively low extinction due to intervening dust and gas. The Hopkins Ultraviolet Telescope (HUT) observed the UV resonance
lines of He~II $\lambda$1640,  C~IV $\lambda\lambda$1548, 1550, N~V $\lambda\lambda$1238,1243, and O~VI $\lambda$1032, 1038 (Raymond et~al. 1995) emitted from the Balmer dominated filament in the NW quadrant. The line showed Doppler broadening consistent with that of the H $\alpha$ broad
component observed in the optical, indicating insignificant ion-ion equilibration. Laming et~al.
(1996) were also able to infer the degree of electron-ion equilibration at the shock.

\begin{figure*}
  \begin{center}
  \includegraphics[width=0.75\textwidth,angle=90]{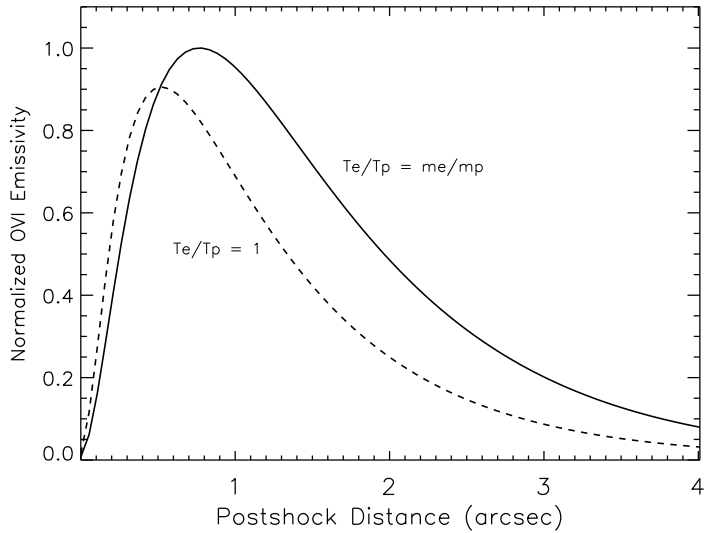}
  \end{center}
\vskip -0.2in
\caption{The spatial distribution of the O~VI(1032+1038) emissivity behind a 350 \kms\, planar shock at a distance of 500 pc (parameters
appropriate to the fastest Balmer-dominated shocks in the Cygnus Loop), shown for
the cases of full, instant equilibration at the shock front (\tetp\,=1) and minimal equilibration (\tetp\,=\,\memp).  Analysis and
modeling of the H$\alpha$ profiles, as well as the O~VI and X-ray emission indicate that \tetp\,$\approx$\,1 for the Cygnus Loop.  }
\label{fig:ovispatial}       
\end{figure*}

He~II $\lambda$1640, by virtue of its relatively high excitation potential $\sim$ 48 eV), is
excited only by electrons, and its intensity is therefore directly related to the electron
temperature. C~IV $\lambda\lambda$1548, 1550, N~V $\lambda\lambda$1238,1243, and O~VI $\lambda$1032, 1038, by contrast, have much lower excitation potentials of $\sim 8, 10$ and 12 eV, so although these ionization states are established by electrons, the line emission in these
transitions can also be excited by impacts with hot protons and $\alpha$ particles, and the intensity ratio of He~II $\lambda$1640 to C~IV $\lambda\lambda$1548, 1550, N~V $\lambda\lambda$1238,1243, and O~VI $\lambda$1032, 1038 can be sensitive to the post-shock
electron-proton temperature equilibration.  The spatial distribution of the UV resonance line emission, when spatially resolved,
provides additional constraints on the degree of electron-ion equilibration at the shock front.  For shocks slower than 1500 \kms,
$T_e\,=\,T_p$ at the shock front results in both a more rapid rise and higher maximum in emissivity of the UV resonance lines with
distance behind the shock (see Figure~\ref{fig:ovispatial}).

Laming et~al. (1996) calculated impact excitation cross sections for protons and $\alpha$ particles colliding with Li-like ions, 
using a partial wave expansion with the Coulomb-Bethe approximation, and applying a unitarization procedure following Seaton (1964). 
They found a degree of equilibration of order $T_e/T_p\sim 0.05$ or less, which implied for a 2250 km s$^{-1}$ shock an electron 
temperature immediately postshock of $< 5\times 10^6$ K, in very good
agreement (and actually predating) the optical results discussed above in subsection 4.1.

The ion-ion equilibration in SN 1006 was revisited by Korreck et~al.(2004), using higher spectral
resolution FUSE data comprising O~VI $\lambda$1032, 1038 and the broad
Ly~$\beta \lambda$1025 emission lines. They found a slightly broader line profile in Ly $\beta$,
implying less than mass-proportional heating and possibly a small degree of ion-ion equilibration.

SN 1987A represents another SN/SNR in a region of the sky accessible to UV observations. HST COS
observed the He~II $\lambda$1640,  C~IV $\lambda\lambda$1548, 1550, N~V $\lambda\lambda$1238,1243 and N~IV $\lambda\lambda$1486
lines emitted from the reverse shock (France et~al. 2011).
When combined with optical spectroscopy of H $\alpha$,
the \tetp\, ratio at the shock is determined to be in the range 0.14 -- 0.35, significantly
higher than similar ratios coming from Balmer dominated forward shocks. France et~al. (2011)
argued that a different equilibration mechanism is likely at work. Considering the relative
youth of SN 1987A, and the fact that the reverse shock is the origin of the emission, significant
populations of cosmic rays and associated magnetic field amplification are unlikely. In fact, in
the expanding ejecta the magnetic field is likely to be very weak, leading to a very high
Aflv\'en Mach number shock. As will be discussed below in connection with shocks in galaxy clusters, 
electron heating in such a case is likely to be due to acceleration in the cross-shock potential.
The cross-shock potential is effective at heating electrons, and so may explain the higher \tetp\,
in SN 1987A.

The forward shock of SN 1987A has also been observed in X-rays with the grating instruments
on Chandra (e.g. Zhekov et ~al. 2009). In general electron heating well below complete
equilibration is seen, though precise interpretation is difficult because one observation sees 
emission from shocks at a variety of different velocities,
due to irregularities in the density of the surrounding medium.

\subsection{Do Results from the Balmer-Dominated Shocks Apply to Fully Ionized Shocks?  }
\label{sec:4}

The inverse relationship between the temperature equilibration and shock speed is
an interesting result from studies of Balmer-dominated SNRs.  However, the applicability of this result
to both fully ionized shocks and shocks undergoing efficient CR acceleration ($\gtrsim$50\% of their
energy transferred to CRs) remains unsettled. Recently, Vink et al. (2010) used a two-fluid-model for
cosmic rays and thermal gas to simulate the effect of cosmic ray acceleration on the temperature and
ionization structure of fast, non-relativistic shocks.  They found that if 5\% of the
shock energy were to be channeled into cosmic rays (the minimum needed if SNRs are the dominant
source of cosmic rays) then approximately 30\% of the postshock pressure must reside in cosmic rays
(corresponding to a ratio of cosmic ray to total postshock pressure, w, of 0.3).
For w\,=\,0.3, Vink et al. (2010) predicted  a lowering of the average temperature of the postshock
gas to $\sim$70\% of the value when the cosmic ray contribution is ignored.  This is a significant
alteration of the postshock temperature profile, and should result in much more rapid equilibration
of electrons and protons close to the shock.

However, do the effects described above actually occur in SNR shocks?  One of the principle
lines of evidence cited by Vink et al. (2010) in support of this picture was the result found in RCW 86
by Helder et al. (2009).  In that SNR, the broad H$\alpha$ widths of Balmer-dominated filaments were found to be nearly 50\%
smaller than the minimum allowed given their X-ray proper motions. Filaments in the NE of the SNR exhibited
broad H$\alpha$ widths of 1000 \kms, but their apparent X-ray counterparts, which exhibited strong
X-ray synchrotron (nonthermal) emission, exhibited proper motions indicating shock speeds of 3000-6000 \kms.
This result, along with the theoretical prediction that X-ray synchrotron emission requires
shock speeds of at least 2000 \kms\, (Aharonian et al. 1999) was taken by Helder et al. (2009)
and Vink et al. (2010) as evidence for substantial energy
loss (w\,$\sim$\,0.5) from the Balmer-dominated shocks to cosmic rays.  However, this association has now
been refuted by subsequent multi-epoch optical imagery
of the H$\alpha$ filaments, which have failed to show the kind of high proper motions seen in the nonthermal X-ray filaments
(Helder et al. 2012, in preparation).  Instead, they show proper motions consistent with shock speeds
predicted by the broad H$\alpha$ widths without energy loss to cosmic rays($\sim$600-1200 \kms), implying that
w\,$<$\,0.2.
The association between the Balmer-dominated shocks studied spectroscopically by Helder et al. (2009) and
the X-ray filaments was due either to coincidental spatial alignment, or due to sudden deceleration of
the outer shock in RCW 86 during its encounter with the surrounding cavity wall (Williams et al. 2011; Helder
et al. 2012, in preparation).

\begin{figure*}
 \begin{center}
  \includegraphics[width=1.1\textwidth]{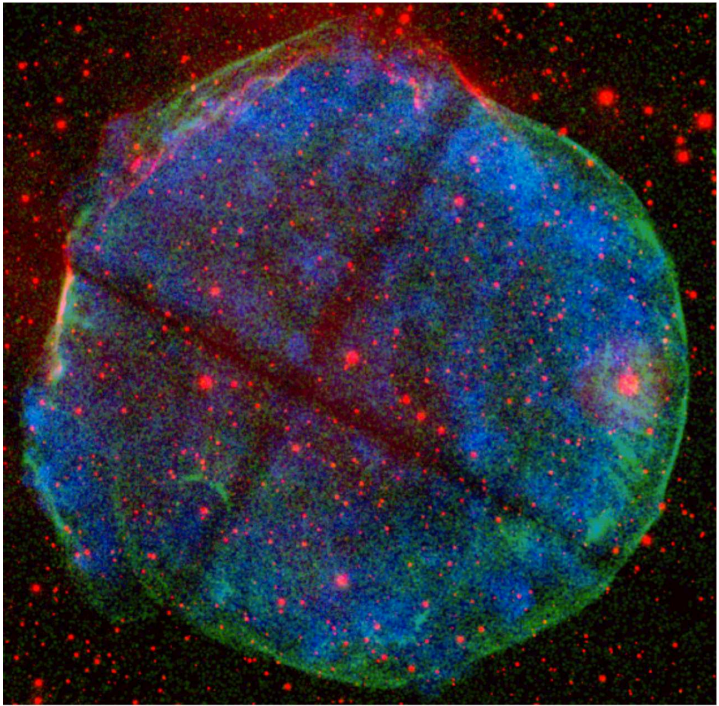}
 \end{center}
\vskip -1.2in
\caption{An H$\alpha$ narrowband image of Tycho's SNR (red; courtesy P. F. Winkler), overlaid onto the
\chan\, 3-8 keV hard X-rays (green) and \chan\, 0.5-3 keV soft X-rays (blue). Both optical and X-ray images
were acquired during the same epoch (2007).  Note the near mutual exclusivity of the Balmer emission and
hard X-ray emission. }
\label{fig:tycho_balmer}       
\end{figure*}

The lack of association between the Balmer-dominated filaments and the non-thermal X-ray
filaments in RCW 86 raises some important questions about the feasibility of using Balmer-dominated shocks to study
electron-proton equilibration in cases where 50\% or more of the thermal energy is diverted to cosmic rays.
In SNRs such as SN 1006 (Koyama et al. 1996; Katsuda et al. 2010a), Tycho's SNR (Warren et al. 2005; Katsuda et al.
2010b), Cas A (Vink \& Laming 2003) and RX J1713.7$-$3946 (Koyama et al. 1997; Slane et al. 1999; Tanaka et al. 2008)
the presence of strong synchrotron X-ray filaments has been interpreted as evidence for highly efficient cosmic ray
acceleration.  The narrowness of the synchrotron filaments most likely reflects the short emitting lifetimes of
the ultra high energy electrons (energies $\sim$10-100 TeV) as they spiral in the postshock magnetic field (Vink \& Laming 2003).
The detection of $\gamma$-ray emission from the shells of SN 1006 (Acero et al. 2010) and RX J1713.7$-$3946 
(Aharonian et al. 2006; Abdo et al. 2011) has shown that cosmic rays are accelerated to energies as high as 100 TeV
in these SNRs.  In all cases  thermal X-ray emission has been exceptionally faint due to
the very low inferred preshock densities ($\lesssim$0.1 cm$^{-3}$), making it more likely that the overall
X-ray emission will be dominated by synchrotron radiation from the most energetic cosmic rays.  SNRs expanding
into such low density media can propagate at the high shock speeds required for cosmic ray acceleration
($\gtrsim$2000 \kms; Aharonian et al. 1999) for a longer time, allowing their structure to become modified
by the back pressure from the cosmic rays.  Investigating the temperature and ionization structure of such
shocks with Balmer line spectroscopy
requires finding Balmer-dominated shocks exhibiting X-ray synchrotron radiation.

All of the known SNRs exhibiting Balmer-dominated shocks have also been imaged at X-ray wavelengths
with \chan\, or XMM, allowing reasonably detailed searches for shocks emitting both H$\alpha$
and synchrotron X-ray emission (the latter producing hard continuum that is dominant at energies of
2 keV and higher). A detailed comparison
for all Balmer-dominated SNRs has not yet been published. However, even a cursory comparison between
the narrowband H$\alpha$ and hard X-ray images of these SNRs shows a distinct {\it anticorrelation}
between shocks emitting in these two bands.  For example, overlaying H$\alpha$ and \chan\, images of Tycho's SNR
acquired during the same epoch (2007) (Figure \ref{fig:tycho_balmer}) shows little or no correlation between the prominent
Balmer-dominated filaments
on the eastern and northeastern edges and the non-thermal X-ray filaments (E$\geq$3 keV) circling the remnant.
The Balmer-dominated filaments (shown in red in Figure \ref{fig:tycho_balmer}) on the eastern side of Tycho's SNR are seen
projected 30$^{\prime\prime}$-1$^{\prime}$ inside the edge of the nonthermal X-ray filaments (marked in green),
an indication that portion of the
shell along the line of sight has significantly decelerated.  The Balmer filaments are seen at the outermost edge
of the thermally emitting X-ray ejecta (marked in blue), but only at locations where little or no nonthermal
X-ray emission is present.  The bright optical filament known as Knot g (at the far left edge of Figure \ref{fig:tycho_balmer})
is the only location where Balmer filaments and
X-ray synchrotron emission appear coincident.  However, upon closer inspection the anticorrelation between the Balmer
line and synchrotron emission can be seen in Knot g as well: the upper half of the filament, where Balmer line emission
is strongest, exhibits minimal synchrotron emission, while the opposite is true in the lower half of the filament.
The enhanced nonthermal emission inside of Knot g may be due to the strong recent deceleration of Knot g, where
the SNR is currently propagating into a strong density gradient at the outermost edge of an H~I cloud (Ghavamian et al. 2000).
The lack of optical/X-ray synchrotron correlation is especially striking given that the Balmer-dominated filaments in Tycho's SNR
have a high enough shock velocity ($\sim$1800-2100 \kms) to accelerate particles to TeV energies.

The anticorrelation between the optical and nonthermal X-ray emission can be observed in other SNRs as well,
including SN 1006, where recently X-ray proper motions have been measured along the entire rim by Katsuda et al. (2012).
As with Tycho's SNR, the locations of the Balmer-dominated filaments and the nonthermal X-ray filaments along the NW rim of SN 1006
are mutually exclusive.  Instead, the Balmer-dominated shocks are closely associated with thermal X-ray filaments
having a proper motion consistent with a shock velocity of 3300$\pm$200$\pm$300 \kms\, (statistical and registrational
uncertainties, respectively) for a distance of 2.2 kpc.  This result is in excellent agreement with the shock velocity
of 2890$\pm$100 \kms\, determined from the broad H$\alpha$ width and broad-to-narrow ratio of the NW filament by Ghavamian et al.
(2002). Such close agreement is a strong indication that little substantial energy has been lost from the
thermal plasma to cosmic ray acceleration, similar to optical proper motion studies from RCW 86 (Helder et al., in preparation).

From the above discussion it appears that Balmer-dominated SNRs, while offering powerful diagnostics of \tetp\,
and \vs, are not useful for investigating equilibration the extreme cases of strongly cosmic-ray modified shocks.
In fact, the very condition allowing for the detection of the Balmer line emission - presence of neutral gas ahead of the shock -
is also responsible for limiting the fraction of shock energy lost to cosmic ray acceleration.  Quantatitive evaluations
of this effect by
Drury et al. (1996) and Reville et~al. (2007) show that when the preshock gas is significantly neutral, Alfv\'en waves driven by the cosmic rays
ahead of the shock are dissipated by ion-neutral damping.  As long as the charge exchange frequency,
$\omega_{cx}$ ($\equiv\,n_{HI} \langle\sigma_{cx} v\rangle$) is
larger than the Alfv\'en wave frequency, $\omega_A$ ($\equiv\,k\,v_A$) the ions and neutrals oscillate
coherently, and ion-neutral damping is not important.  However, when $\omega_{cx}\,<\,\omega_A$ the neutrals are left
behind by the ions in the Alfv\'en wave motion, and during the incoherent oscillation between the two, charge exchange
exerts a drag on the Alfv\'en waves, damping them.
The condition required for Alfv\'en waves to not be strongly damped in the precursor can be written out as
\begin{equation}
n_{HI}\langle \sigma_{cx}v\rangle\,\,>\,\,k\,v_A
\end{equation}
where $v_A\,\equiv\,\frac{B}{(4\pi\,m_i\,n_i)^{1/2}}$ is the Alfv\'en speed of the ions ahead of the shock
and $n_{HI}$ is the preshock neutral density. Given that cosmic rays resonantly scatter off
Alfv\'en waves having Doppler shifted frequencies comparable to their gyrofrequency, and that the cosmic ray gyrofrequency is
related to its energy via $\omega_{cr}\,\equiv\,e\,c\,B / E$, the inequality above can be cast in terms of the energy, $E_{crit}$,
below which a significant fraction of the cosmic ray flux out of the shock is reduced by ion-neutral damping:
\begin{equation}
E_{crit}(TeV)\,=\,0.07\,\frac{B^2_3\,T_4^{\,-0.4}}{x_{HI}\,(1 - x_{HI})^{1/2}\, n^{3/2}}
\end{equation}
where we have set $\langle\sigma_{cx}v\rangle\,\approx$\,8.4$\times$10$^{-9}\, T_4^{\,0.4}$ cm$^{3}$ s$^{-1}$ (Kulsrud \& Cesarsky 1971),
$B_3$ is the preshock magnetic field strength in units of 3 $\mu G$, $n$ and $x_{HI}$ are the total preshock density and neutral fraction, and $T_4$ is the preshock temperature in units of 10$^4$~K.  For Balmer-dominated SNRs, where recent models have required
moderate amplification of the preshock magnetic field ($\Delta B/B\,\sim\,$3-5; Ghavamian et al. 2007) and where the preshock
temperature may exceed 20,000~K (Raymond et al. 2011), $E_{crit}\,\sim\,$4 TeV for the typical case where
$x_{HI}\,=\,$0.5 cm$^{-3}$.  SNRs exhibiting nonthermal X-ray emission are believed to contain cosmic rays with energies
of tens of TeV, so $E_{crit}\,\sim$4 TeV is certainly high enough reduce the effectiveness of Balmer-dominated shocks 
in producing nonthermal X-ray emission. 

However, as noted earlier, a modest back pressure from cosmic rays is required to explain the width of the
H$\alpha$ narrow component line, as well as the low broad-to-narrow ratios seen in some SNRs (Rakowski et al. 2009;
Raymond et al. 2011).   In fact, one model for electron heating in fast collisionless shocks
requires at least some feedback from the cosmic rays in order to explain the moderate heating of electrons
in SNRs, as well as the inverse squared relationship between \tetp\, and \vs\, (Ghavamian et al. 2007, described
in the next section).  Furthermore, as pointed out by Drury et al. (1996), the ion-neutral damping of 
Alfv\'en waves in the precursor is unimportant for cosmic rays which have already exceeded $E_{crit}$.  Since
the acceleration time for cosmic rays shortens considerably with shock speed ($\tau_{acc}\,\approx\,\kappa_{CR}/V_s^2$; 
Malkov \& Drury 2001), the fastest Balmer-dominated shocks are more likely to have accelerated particles beyond
$E_{crit}$ and hence will begin to exhibit nonthermal X-ray emission and cosmic-ray modified shock structure.  
A good example is the aforementioned SNR 0509$-$67.5, where
the shock speeds exceed 5000 \kms\, (Helder, Kosenko \& Vink 2010) and nonthermal X-ray emission from cosmic ray
accelerated electrons is detected from the forward shock.  The forward shocks in more evolved Balmer-dominated SNRs 
(such as SN 1006 and Tycho's SNR) will have swept up more mass and slowed down to speeds $\lesssim$2000 \kms, by
which point $\tau_{acc}$ will have lengthened and the shocks will be less cosmic-ray dominated.

\section{Models for Electron Heating in SNRs}
\label{sec:5}

Given the lack of in situ measurements of the particle distributions in SNRs, the electron heating mechanisms
in these shocks have been studied primarily via numerical methods.  On one hand, a number of studies have focused on electron
heating in relativistic shocks, with the aim of modeling high energy emission from gamma-ray bursts
(e.g., Gedalin et al. 2008; Sironi \& Spitkovsky 2011).  These shocks are in a different area of parameter space than the SNR shocks
discussed here, and the physics governing the electron heating in relativistic shocks is substantially different.  At the very high Alfv\'enic Mach numbers
characteristic of gamma-ray bursts, the shock transition becomes very thin (less than an electron gyroradius).  Electrons in this case may be
once again accelerated by the cross shock potential, similar to the very low Mach number case. On the other hand,
a number of other studies consider non-relativistic shocks relevant to SNRs ($\lesssim$0.01c), where accelerated particles 
such as cosmic rays or solar energetic particles (SEPs) may play an important
role in establishing their shock structure.  These studies have sought to
identify plasma waves capable of boosting electrons to mildly relativistic energies (e.g., Amano \& Hoshino
2010; Riquelme \& Spitkovsky 2011), with the objective
of understanding how electrons are injected into the cosmic ray acceleration process.  This is a different
(though related) question from what we consider here, namely how electrons are promptly heated to temperatures $\sim$5$\times$10$^6$~K
at the shock front (Ghavamian et al. 2007; Rakowski et al. 2008).  This limits our consideration of the work
done so far to two broad scenarios of electron heating in fast, non-relativistic collisionless shocks.
One scenario is based on lower hybrid wave heating in the cosmic ray precursor (Laming 2000; Ghavamian
et al. 2007; Rakowski et al. 2008) while the other is based on counterstreaming instabilities ahead of the
shock (e.g., the Buneman instability, Cargill \& Papadapoulos (1988), Matsukiyo 2010; Dieckmann et al. 2012).
We discuss these mechanisms below in turn.

\subsection{Lower Hybrid Wave Heating}

The most significant result of the Balmer-dominated shock studies, the inverse squared relation between
\tetp\, and \vs, places a useful constraint on the range of plausible equilibration mechanisms at the shock
front.  The simplest way to obtain \tetp\,$\propto\,V^{-2}_{s}$ is to set $\Delta\,T_e\,\approx\,const.$
at shock speeds of 400 \kms\, and higher, while allowing $T_p$ to rise according to the Rankine-Hugoniot
jump conditions, $k \Delta\,T_p\,\approx\,\frac{3}{16}m_p V^2_{s}$.  The requirement that \tetp\,=\,1
at \vs\,=\,400 \kms\, gives $\Delta\,T_e\,\approx\,$0.3 keV for \vs\,$\geq$400 \kms, independent of
shock velocity (Ghavamian et al. 2007).  Although there may be marginal evidence of a departure from this
relation at shock speeds exceeding 2000 \kms\, (van Adelsberg et al. 2008), a velocity-independent heating
of electrons in SNR shocks is an important clue to the nature of plasma heating processes in fast
collisionless shocks. It suggests that plasma processes ahead of the shock front are an important
(if not dominant) source of electron heating in SNRs (Ghavamian et al. 2007; Rakowski et al. 2009).

As mentioned earlier, strong interstellar shocks are expected to form a precursor where cosmic rays
crossing upstream give rise to Alfv\'en waves and turbulence (Blandford \&
Eichler 1987; Jones \& Ellison 1991), compressing and pre-heating the gas before it enters the shock.
As long as the shock is strong ($v_{downstream}/V_{s}\,\approx\,$1/4) and cosmic ray pressure does not
dominate the postshock pressure ($\Delta$B/B does not greatly exceed unity, with $\lesssim$20\% of the
postshock pressure provided by
cosmic rays) the thermal heating within the precursor does not depend strongly on shock velocity.  The
limited range of narrow component H$\alpha$ widths observed in Balmer-dominated SNRs over a wide range in
shock speeds (Sollerman et al. 2003; Raymond et al. 2011) is consistent with the relative insensitivity of
the preshock heating to shock speed.  Since the widening of the H$\alpha$ narrow component line is now believed
to arise in a precursor where the gas is heated by the damping of cosmic-ray driven waves (Wagner et al. 2008;
Raymond et al. 2011; Morlino et al. 2012), 
it stands to reason that perhaps the physical processes generating a constant electron heating with
shock speed ($\Delta\,T_e\,\approx\,$0.3 keV) also originate within the cosmic ray precursor.

The above argument was used by Ghavamian et al. (2007) and Rakowski et al. (2008) to advocate for a heating model
where lower hybrid waves within the cosmic ray precursor preheat electrons to a constant temperature before
they enter the shock front.  This model was based on the conception of McClements et al. (1997), who suggested
that lower hybrid waves driven by the reflected population of nonthermal ions could generate lower hybrid
waves ahead of the shock, pre-heating electrons and injecting them into the cosmic ray acceleration process.
The condition for generating such waves is that the shock be quasi-perpendicular, and that the reflected
ions form a beam-like (gyrotropic) distibution.  A similar scenario was suggested by Ghavamian et al. (2007)
and Rakowski et al. (2008), but with one crucial difference: the reflected particles considered are ultra-relativistic
cosmic rays rather than suprathermal ions.  The lower hybrid waves are electrostatic ion waves which propagate
perpendicular to the magnetic field and whose frequency is the geometric mean of the electron and ion geofrequencies,
$\omega_{LH}\,=\,(\Omega_e\,\Omega_i)^{1/2}$.  The group velocity of these waves is directed primarily along
the magnetic field lines ($k^2_{||}/k^2_{\perp}\,=\,m_e/m_p$; Laming 2001) and the waves can simultaneously resonate with ions
moving across the field lines and electrons moving along the field lines.  Although the growth rate of lower
hybrid waves is generally small (Rakowski et al. 2008), their group velocity perpendicular to the magnetic field (and hence the
shock front) can be on the order of the shock velocity ($\partial \omega/\partial k_{\perp}\,\approx\,$\vs).  This
allows the lower hybrid waves to remain in contact with the shock for long periods of time, attaining high
intensities capable of effectively heating the electrons (McClements et al. 1997; Ghavamian et al. 2007).

In the case of cosmic rays, the time spent by the electrons in the precursor is
$t\,\sim\,\kappa_{CR}/v^2_{sh}$.  The kinetic energy acquired by the electrons in the precursor is
$\Delta\,E_e\,\propto\,D_{||\,||}\,t$, where $D_{||\,||}$ is the momentum diffusion coefficient of electrons
(Ghavamian et al. 2007).  For lower hybrid wave turbulence, $D_{||\,||}\,\propto\,V_{s}^2$ (Karney 1978; Ghavamian
et al. 2007; Rakowski et al. 2008), so that $\Delta\,E_e\,\approx\,\frac{1}{16}\,\left(\frac{m_e}{m_p}\right)^{1/2}\,
m_e \Omega_e\,\kappa_{CR}\,\propto\,B\,\kappa_{CR}\,\sim\,const$, as needed to account for the inverse squared
relationship between equilibration and shock speed.  Note that under the assumption that nonlinear amplification
of the preshock magnetic field is not too strong ($\Delta\,B/B\,\sim\,1$), $\kappa_{CR}$ is that of Bohm diffusion,
which scales as 1/B, so that $\Delta\,E_e$ is also approximately independent of B.

During the past decade more refined models of cosmic ray acceleration have shown that a non-resonant mode
of Alfv\'en waves, having a higher growth rate than the previously considered resonant mode (Skilling 1975), can be excited by cosmic
rays in the precursor (Bell \& Lucek 2001, Bell 2004, 2005).   Unlike for the resonant case, the non-resonant
amplification allows for $\Delta\,B/B\,\gg\,1$, driving preshock magnetic fields to values as high as
1 mG (Vink \& Laming 2003; Berezhko et al. 2003; Bamba et al. 2005;
Ballet 2006).  Such magnetic fields are hundreds of times stronger than the canonical preshock magnetic field
of 3 $\mu G$ and high enough to account for the observed narrowness of
X-ray synchrotron-emitting rims in such SNRs as SN 1006 (assuming the narrowness is due to rapid cooling
of high energy electrons behind the shock; see Ballet 2006 and Morlino et al. 2012).  Additional studies have
suggested that non-resonant amplification may dominate early in the life of the SNR, while resonant amplification
may take over during the Sedov-Taylor stage of evolution (Amato \& Blasi 2009; Schure et al. 2012),
though in either case, $\Delta\,B/B\,>$10 is readily attained.  Such a strong magnetic field
effectively reduces the acceleration time for particles, and is very well suited for explaining how
cosmic rays can reach the knee in the cosmic ray spectrum near 10$^{15}$ eV (Bell \& Lucek 2001; Eriksen et al. 2011).

An important factor influencing the effectiveness of lower hybrid wave heating of electrons is the orientation
of the preshock magnetic field relative to the shock front.  Lower hybrid wave heating is only effective in
perpendicular shocks (Vink \& Laming 2003, Ghavamian et al. 2007;  Rakowski et al. 2008).  Given their spherical
global geometry, SNR blast waves generally propagate at a range of angles to the interstellar magnetic
field.  X-ray observations and models of such SNRs as SN 1006 (Orlando et al. 2007; Petruk et al. 2008) have indicated
that perpendicular shocks are far more effective at accelerating cosmic rays than parallel shocks.  Although
the detailed implications of such differences have not yet been worked out for the lower hybrid wave heating model,
Rakowski et al. (2008) argue that even for quasi-parallel shock geometries the cosmic ray current driving the
nonresonant Alfv\'en waves will generate a significant perpendicular magnetic field ahead of the shock (such a
possibility has also been inferred from numerical simulations; Riquelme \& Spitkovsky 2011).  This
would allow lower hybrid wave growth to overtake modified Alfv\'en wave growth for arbitrary orientations of the far
upstream magnetic field, and allow for a more ubiquitous role for lower hybrid wave heating of electrons.

The amplification of the preshock magnetic field well beyond its far upstream value introduces an interesting
possibility: effective lowering of the Alfv\'enic (and hence magnetosonic) Mach number of the shock.  For the
Balmer-dominated shocks, where analysis of the optical spectra has shown that at best only a moderate fraction
($\lesssim$20\%) of the shock energy has likely been channeled into cosmic rays, the widening of the H$\alpha$
narrow component has been interpreted as nonthermal broadening caused by the lowest frequency waves in the
precursor (Ghavamian et al. 2007; though see Raymond et al (2011) for a thermal intepretation) .  To explain
the 30-50 \kms\, widths of the H$\alpha$ narrow component, the preshock magnetic field must be enhanced by
a factor of a few.  For the non-resonant Alfv\'en waves in the Bell (2004) mechanism, the magnetic field energy
density immediately behind the shock is given by (Schure et al. 2012)
\begin{equation}
\frac{B^2}{4\pi}\,\approx\,\frac{1}{4}\,\phi^2\,\rho\,v^2_{sh}
\end{equation}
where $\phi\,\equiv\,P_{CR}/\rho\,v^2_{sh}$ is the fraction of the shock ram pressure channeled into cosmic
rays.  Solving this expression for B gives $B(\mu G)\,\approx\,228.7\,\phi\,n^{1/2}\,V_{1000}$, where
$V_{1000}$ is the shock speed in units of 1000 \kms.   For $\phi\,\sim\,$0.1-0.2, $n\,\sim\,$1 cm$^{-3}$,
a postshock compression factor of 4 and Balmer-dominated shock speeds $\sim$2000 \kms, this gives
$\Delta\,B/B\,\sim$4-10 ahead of the shock.  Correspondingly, $v_A$ can increase by nearly the same factor,
so that $M_A$ can be reduced by as much as an order of magnitude.  Treumann \& Jaroschek (2008) describe the
physical picture in this case as that of the shock having to prevent an increasing number of ions from
crossing the shock jump by deflecting an increasing number of them at higher and higher Mach numbers.  This
deflection is necessary so that the ability of the shock to dissipate the inflowing energy is not overwhelmed.  By
deflecting these ions back upstream into a precursor, the net inflow of momentum and energy density into
the shock is reduced, reducing the net difference in velocity between the inflowing and outflowing ions.  This
effectively reduces the Mach number in the frame of the upstream medium.

Note that the ions in the precursor are only mildly compressed (Wagner et al. 2008; Morlino et al. 2012), which
only weakly counteracts the rise in B.  In addition, $v_A$ only scales as $n^{-1/2}$, but scales directly as B.
The result is that {\it given the compelling evidence for enhanced preshock magnetic fields in SNRs shocks, the
Mach numbers of these shocks may be overestimated by as much as an order of magnitude.}  As we describe in Section
\ref{sec:5}, a unified description of solar wind and SNR shocks, where the physics of electron-ion temperature equilibration
occurs over a similar range in Mach numbers and involves a similar range of physical processes, may be possible.

\subsection{Plasma Wave Heating from the Buneman Instability}

Similar to the lower hybrid wave model, the Buneman instability-driven wave model focuses on
the region immediately ahead of a quasi-perpendicular shock.  However, unlike the lower hybrid wave
model, the Buneman instability models
consider the reflected nonthermal ion distribution, rather than ultrarelativistic cosmic rays.  In the latter model,
$\sim$20\% of the ions are reflected
backstream against the incoming electron and ion plasma (Papadapoulos 1988; Cargill \& Papadapoulos 1988).  Here
the upstream plasma is not electrically neutral due to the positive charge of the reflected ion
distribution.  In such cases, a drift is induced between the electrons and ions.   The microinstabilities excited by
this configuration depend upon the relative
size of the electron thermal speed relative to the electron-ion drift velocity.
The Buneman instability occurs when the drift velocity of the reflected
ions relative to the upstream electrons exceeds the thermal
speed of the upstream electrons ($2 v_{s}\,>\,(kT_e/m_e)^{1/2}$)(Cargill \& Papadapoulos 1988), a condition
which occurs for very high Mach number ($M_A\,\gtrsim$50) shocks.  If the reflected
ion current upstream is strong enough, the electron current generated to counteract it may produce a large enough
drift between the preshock ions and electrons to cause a secondary Buneman instability when
the ion speed exceeds the electron thermal speed (Dieckmann et al. 2012).  The Buneman instability generates electrostatic plasma waves
which damp by rapidly heating the preshock electrons to $k\Delta\,T_e\,\approx\,2 m_e v^2_{s}\,\approx\,0.01\,v^2_{1000}$ keV,
where $v_{1000}$ is the shock speed in units of 1000 \kms. until their thermal speed matches the electron-ion
drift speed, at which point the instability saturates.  The rapid heating of the electrons perpendicular to the
magnetic field results in $T_e/T_i\,\gg\,$1 and
makes it possible for an ion acoustic instability to occur between the preshock electrons and either the reflected ions
or preshock ions (Cargill \& Papdapoulos 1988).  The waves generated by the ion acoustic instability can then
transfer a substantial fraction of the shock energy (tens of percent) into electron thermal energy.  This process
occurs over a length scale of $v_{s}/\Omega_i$ (as opposed to $\kappa_{CR}/v_{s}$ for the cosmic ray precursor),
resulting in a \tetp\,$\approx$\,0.2, independent of shock speed.  {\it This is in strong disagreement with the equilibrations
obtained for the Balmer-dominated shocks.}

A number of other electron heating mechanisms, such as the modified two-stream instability and electron-cyclotron drift
instability,  have been proposed for collisionless shocks based on results
from particle in cell (PIC) simulations (Umeda et al. 2012a, 2012b).  A unified picture proposed by Matsukiyo (2010) suggests that
electrons can be strongly energized at low Mach numbers ($M_A\,\leq\,$10) via a modified two-stream instability, where
the velocities of the reflected/incoming ions and the electrons are lower than the thermal speed of the electrons
and the electrons are able to damp out the Buneman instability.  In this case, obliquely propagating whistler mode
waves are excited, having frequencies between the electron cyclotron frequency and the lower hybrid wave frequency.
When the electron-ion drift velocity and electron thermal speed become nearly equal, the electron-cyclotron
drift instability becomes important (Umeda et al. 2012a), exciting waves with frequencies that are multiples of
the electron cyclotron frequency.  At higher Mach numbers the electron thermal speed is lower than that
of the ions, and the Buneman instability/ion acoustic wave process descrbied earlier is predicted to take over.

The amount of electron heating predicted by the Buneman instability/ion acoustic wave model scales as $M^2_A$
(Cargill \& Papadapoulos 1988; Matsukiyo 2010), so that for shocks in the 2000 \kms\,-\,10,000 \kms\, range, $\Delta\,E_e\,\sim\,
$2-50 keV.  This is clearly at odds with $\Delta\,E_e\,=\,$0.3 keV observed between 400 \kms\, and 2000 \kms\,
for Balmer-dominated shocks.  One explanation for this discrepancy is that growth of the Buneman-like and
two-stream instabilities described above requires that the reflected ions form a distribution function with a
positive gradient at some velocity (Laming 2000).  This distribution forms when specularly reflected ions have a mostly
monoenergetic, beamlike configuration.  At the low Mach numbers in the solar wind ($\lesssim$10), where the shock
structure is laminar, the reflected ions closely resemble a monoenergetic beam.  However, at the higher Mach numbers,
where the shock front is more turbulent and disordered, the reflected ions are likely to have a greater spread in energy and
are probably less beamlike (Laming 2000). This would lead to suppression of Buneman-like instabilities.  However, this
line of reasoning is still speculative, and the real explanation for the lack of agreement between the observed \tetp\,
and those predicted by models in this section remains to be explored.  Cosmic-ray driven processes may ultimately
provide a better explanation for electron heating at SNR shocks than those involving reflected suprathermal ions.

The cross shock potential arises from the charge separation produced by the different gyroradii
for ions and electrons as they cross the shock transition. It accelerates electrons into the
post shock layer, and can be a means of electron heating at subcritical shocks with an approximately laminar structure. At supercritical shocks, the time dependence and non-locality
introduced reduces the degree of electron heating. However at sufficiently Alfv\'en Mach number (where the shock transition becomes thin 
(on the order of the electron convective gyroradius or inertial length),
significant electron heating may again occur. In the absence of magnetic
field amplification by cosmic rays, this might be expected to happen at SNR shocks. However it is
much more likely in environments where the plasma beta is low, such as galaxy clusters.  It may also occur in cases 
where a significant population of cosmic rays is unlikely due to the low age of the shock, as
in gamma-ray bursts.

\section{Constraints From Solar Wind Studies}
\label{sec:6}

From the beginning, a detailed study of the fastest collisionless shocks has been
hampered by one inherent limitation: they occur in objects which are too remote for in situ study.
Although some collisionless shocks in our solar system reach Alfv\'enic Mach numbers as high as 30
(such as those around Saturn; Achilleos et al. 2006; Masters et al. 2011), there are no physical phenomena in our solar system energetic
enough to produce the type of shocks seen in SNRs (Mach numbers $\sim$100 or more if no preshock magnetic
field enhancement in the CR is assumed).
In addition, the range of plasma betas attainable in the solar system is larger than the range of plasma betas
attained in the interstellar medium.

Another fundamental difference between solar wind and SNR shocks is the fact that the former are short-lived
phenomena confined to small spatial scales (millions of km) in a curved (bow shock) geometry, whereas the latter
are sustained for thousands of years, on spatial scales of parsecs, often well-described by a planar geometry.
This results in an irreducible difference between the two types of shocks: particles crossing back and forth
between upstream and downstream can remain in contact with SNR shocks for long periods of time, allowing
accelerated CRs to acquire much more energy in SNR shocks than in solar wind shocks.  This potentially allows
the CRs to create shock precursors with properties needed to heat electrons and influence \tetp.


The heating of electrons at the Earth's and other planetary bow shocks has been
the subject of much theoretical and observational work. Typical features of the
electron temperature change, $\Delta T_e$, observed at the bow shock as observed
by \citet{schwartz88} include (a) an approximate relationship between heating
and the incident solar wind energy $\Delta T_e \propto U^2$, where $U$ is the
component of the solar wind's velocity incident upon the shock, and (b) a
relationship between the change in temperature normalized by the incident energy
and the fast magnetosonic Mach number, $\Delta T_e / (m_p U^2/2) \propto
M_{ms}^{-1}$. A similar approximate relationship holds between the normalized
electron temperature change and the Alfv\'en Mach number $M_A$, especially for
shocks with a low plasma $\beta$, which is the ratio between thermal and
magnetic pressures. Recent work by \citet{masters11} shows that this
relationship with $M_A$ holds well at Saturn's bow shock. This is particularly
interesting as $M_A$ at Saturn is often much larger than at Earth.

\begin{figure}
  \includegraphics[angle=90, width=5in]{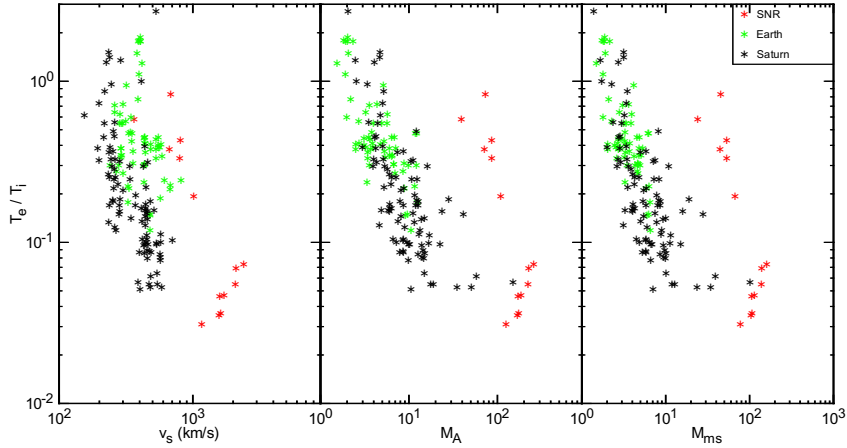}
    \vskip -.5in
  \caption{Collected electron-ion equilibration data from both the solar wind bow shocks
and supernova remnant shocks.  \tetp\, is plotted versus shock speed (left), Alfv\'enic
Mach number (center) and magnetosonic Mach number (right). Green symbols show data from crossings of Earth's
bow shock (\citet{schwartz88}), while the black symbols show data from crossings of
Saturn's bow shock (\citet{masters11}).   Shock
speeds for the Saturnian bow shock are based on a solar wind model
and an assumed shock speed with respect to the spacecraft of 100
\kms, and ion temperatures are based on electron distribution
measurements and the application of the Rankine-Hugoniot conditions
(see Masters et al. (2011) for a full discussion of shock parameter
derivations at Saturn's bow shock).  Red symbols show data acquired from Balmer-dominated SNR shocks
(van Adelsberg et al. 2008), and assume $v_A$\,=\,9 \kms, $c_s$\,=\,11 \kms.  
}
    \label{fig:combined}
\end{figure}

In Figure \ref{fig:combined} we plot the ratio of electron and ion temperatures downstream of
the shock against the magnetosonic, and Alfv\'en Mach numbers, as well as the
upstream flow velocity relative to the shock. The data in these figures are
taken from bow shock crossings of the ISEE spacecraft, and is a subset of those
listed in \citet{schwartz88}, consisting of 61 crossings for which all the
necessary data is available.

It is well known that quantities other than those displayed here may be more
appropriate, specifically the change in electron temperature over the change in
total temperature $\Delta T_e / \Delta (T_e + T_i)$ or even $\Delta T_e / \Delta
T_i$ are better correlated with inverse Mach numbers than $T_e / T_i$
\citep{schwartz88}. Nevertheless, we use the latter quantity here as it enables
a comparison with data from extra-solar system and outer planetary shocks at
where less data are available. Furthermore, the approximate inverse dependence
upon $M_{ms}$, $M_A$, and $V_s$ is still quite apparent in this data. It is
interesting to note that the relationship with the Mach numbers is much more
favourable than the dependence upon $V_s$, indicating that the Mach number is
the more relevant quantity for organizing the relationship between \tetp\, and
shock strength.

Many mechanisms may be involved in the heating of electrons in solar system
shocks. Proposed mechanisms involve acceleration of electrons by a cross-shock
potential \citep{goodrichScudder84, scudder1_86, scudder3_86}, wave turbulence
\citep{galeev76}, microinstabilities \citep{wu84}, and electron trajectory
scattering \citep{Balikhin93}. The existence of a cross-shock potential may be
deduced from the generalized Ohm's law in which a gradient in electron thermal
pressure gives rise to an electric field. Examining the energetics of electrons
crossing the shock may simplified by working in the de Hoffmann-Teller frame of
reference \citep{dehoffmannTeller50}, defined as the frame in which the shock is
at rest, and in which the magnetic field and plasma flow velocity are
(anti-)parallel. In this case the the electric field is dominated by that
generated by the electron pressure gradient, and the work done on electrons
crossing the shock is determined by the cross-shock potential. Additional
mechanisms are required to scatter electrons to pitch angles that are more
perpendicular, and to flatten the distribution so that empty regions of phase
space are filled.  This results 
in a distribution whose temperature is controlled to a large extent by the de
Hoffmann-Teller frame cross-shock potential and downstream density. In addition
to direct measurement of the electric fields within the shock,
\citep{baleMozer07, dimmock11}, comparison of upstream and downstream electron
phase space distributions have shown that these are consistent with electron
acceleration by a cross-shock potential in the de Hoffmann-Teller frame
\citep{lefebvre07}.

\section{Connecting the Solar Wind Results to those in SNRs}
\label{sec:7}

Figure \ref{fig:combined} may indicate that similar mechanism(s) heat the electrons in
solar wind and in SNR shocks.  {\it This is especially appealing when we remember that $M_A$ in SNRs may be overestimated
due to preshock amplification of magnetic field by cosmic rays.} In their study of the terrestrial 
bow shock and interplanetary shocks, Schwartz et al (1988) found that the electron-ion temperature equilibration
organizes best by $T_e/T_i\propto 1/M_A$. Given the difficulty in determining the Mach numbers of SNR shocks,
the equilibration dependence on shock strength has been characterized via the shock speed instead, and
found to obey $T_e/T_i\propto$ 1/\vs$^{2}$\, (\vs\, is
much more accurately known than the Mach numbers). Subsection 4.1 outlined a model for SNR
electron heating, where the cosmic ray diffusion coefficient $\kappa _{CR}$ is assumed
independent of \vs. From quasi-linear theory (Blandford and Eichler 1987),
\begin{equation}
\kappa _{CR} = {p^2c^2v\over 3\pi e^2U}
\end{equation}
where $U$ is the energy density of turbulence ($\equiv\,\langle\Delta\,B^2\rangle/8\pi$) and $v$ is the
cosmic ray velocity.  For resonant amplification, we evaluate
$U$ at $k_{\Vert} = \Omega /v_{\Vert}$ , where $\Omega$ is the gyroradius and  
$v_{\Vert}$ the parallel component of the cosmic ray velocity.  For relativistic cosmic rays, 
where $v=c$, this results in $\kappa _{CR}\propto p^2/U$, assumed constant with
$V_s$. However, for nonrelativistic suprathermal particles, $v$ will most likely be proportional to
the shock velocity \vs, which with the same assumptions leads to $T_e/T_i\,\propto$\,1/\vs\, for
solar wind shocks (as opposed to $1/V_s^2$ in SNRs). This argument is admittedly loose, and 
should not be be viewed as much more than a hypothesis to motivate further work.

In our arguments above we have made considerable assumptions about $\kappa_{CR}$. The most obvious one is that
$\kappa _{CR}$ as written above applies to parallel shocks, whereas we are most likely dealing
with quasi-perpendicular cases. This may reduce the difference anticipated between solar wind
and SMR shocks, depending on the turbulence spectrum (summarized in Appendix A of Rakowski et al. 2008).

The degree of cosmic ray magnetic field amplification at SNR shocks required to bring SNR
data points in Figure~\ref{fig:combined} into alignment with solar wind data points is approximately an order of magnitude
or less at the highest velocities considered.  Given the degree of magnetic field amplification expected from cosmic
ray acceleration, this appears to be highly plausible. In the case
of saturated nonresonant instability (Bell 2004, 2005) by resonant scattering
(Luo \& Melrose 2009), $\langle\Delta B\rangle^2/B^2\sim 10 - 100$ is expected. In the case of nonresonant
saturation, higher values, but strongly dependent on $k$, are predicted. Saturation
by electron heating (i.e. the $M_A$ where growth of lower hybrid waves becomes greater than
the growth rate of magnetic field) leads to similar magnetic field enhancement, with $\Delta B^2/B\sim 200$ (Rakowski et al. 2008). 
Such magnetic field amplification is less likely at solar wind shocks. The suprathermal
particle densities are lower in solar wind shocks, and the ambient magnetic fields are higher, much closer
to where the nonresonant instability would saturate (if not already beyond it).

\section{Observational Constraints from Galaxy Cluster Shocks}
\label{sec:8}

Collisionless shocks occur over a vast range of length scales, with those in galaxy clusters
being among the largest.  While the shock speeds in the galaxy cluster shocks are similar to
those in supernova remnants (up to 4000 \kms; Markevitch 2005; Markevitch \& Vikhlinin 2007; Russell et al. 2012), 
they occur in environments that
are substantially different from both the solar wind and the ISM.  These differences can be
encapsulated via the plasma beta, defined as the ratio of the thermal pressure to the magnetic
pressure of a plasma ($\beta\,\equiv\,n\,k\,T / (B^2 / 8\pi)$).
Utilizing solar wind parameters listed by Bruno \& Carbone (2005), this
ratio ranges from around unity at 1 AU under fast solar wind conditions (wind velocity $\sim$900 \kms) to
around 20 for the quiescent wind (wind velocity $\sim$300 \kms).  The plasma $\beta$ of the ISM
is close to that of the fast solar wind, $\beta_{ISM}\,\sim$1-4 (assuming a B\,=\,3\,$\mu$G, n\,=\, 1 cm$^{-3}$
and T\,=\,10$^4$~K).
On the other hand, the electron temperature of the intracluster medium (ICM) ranges from ∼ 10$^7$K - 10$^8$K (1-10 keV),
with number densities steeply declining from $\sim$ 10$^{−2}$ cm$^{−3}$ near the cluster centers
to $\sim$10$^{−4}$ cm$^{−3}$ at the outer edges.  The corresponding sound speed is close to 1000 \kms,
nearly two orders of magnitude higher than in the general ISM.  The magnetic fields measured in galaxy clusters
are actually close that of the ISM, typically on the order of a microGauss (Carilli \& Taylor 2002).
Therefore, $v_A\,\sim\,$50 \kms\, in galaxy clusters, so that $\beta_e\,\gg\,$1 and the magnetic
field pressure has negligible contribution to the dynamics of shocks in galaxy clusters.  This puts galaxy cluster
shocks in a different region of parameter space than solar wind and SNR shocks.

Clusters such as 1E0657$-$56 (the 'Bullet Cluster') and A520 show strongly enhanced X-ray emission
from collisionless shocks, formed during major mergers when gas from one cluster plunges through gas from the other
(Markevitch et al. 2005; Markevitch \& Vikhlinin 2007; Russell et al. 2012).   Shocks moving mostly along the
plane of the sky have a favorable viewing geometry and appear as giant curved structures hundreds of kiloparsecs
in length.  The large clumps of infalling gas drive bow shocks into the cluster gas, which has already been
heated to at least 1 keV, and produces thermal Bremsstrahlung emission peaking close to that
energy.  This is another major difference between collisionless shocks in the ISM and those
in the ICM. While the Alfv\'enic and magnetosonic Mach numbers of SNR shocks are very
difficult to determine due to the lack of available observational constraints on magnetohydrodynamic
quantities upstream (such as $T$, $B$ and $n$), those in galaxy clusters can readily be measured by spectral analysis of X-ray emission
upstream.   Comparison of this emission to that of the enhanced postshock region gives the density contrast
between the downstream and upstream gas (i.e., $n_2/n_1$).  This density contrast yields the sonic Mach number, $M$, via
solving the Rankine-Hugoniot jump conditions (Russell et al. 2012):
\begin{equation}
M\,=\,\left( \frac{2\,n_2/n_1}{\gamma\,+\,1\,-\,\frac{n_2}{n_1}(\gamma-1)} \right)^{1/2}
\end{equation}
where $\gamma$ is the ratio of specific heats of the cluster gas.  Measurement of these jumps from X-ray 
observations have yielded $M\,\sim\,$1.5-3 for the Bullet Cluster (Markevitch et al. 2005),
Abell 520 and Abell 2146 (Russell et al. 2012).  Using these estimated Mach numbers, the shock velocity itself
can be calculated via \vs\,=\,$M\,c_s$, where $c_s$ is the upstream sound speed as inferred from the
X-ray spectra.  This yields shock speeds ranging between 2500 \kms\, and 4000 \kms, similar to the fastest
known Balmer-dominated shocks.  However, fits to the X-ray spectra behind these shocks show prompt electron-ion
equilibration at the shock front, consistent with \tetp\,=\,1 (Markevitch et al. 2005; Markevitch \& Vikhlinin 2007),
despite the extremely high shock speeds involved.  This result can only be reconciled with equilibration measurements
from the solar wind and SNRs if the equilibration depends on the Mach number, rather than \vs.  Furthermore, given the low
Mach numbers found in the galaxy clusters, it is plausible that the shock transitions in these cases are laminar (as opposed
to turbulent like the SNR and fastest solar wind shocks), with electron heating occurring efficiently at the shock front
via the same type of cross-shock potential as seen in the slowest solar wind and slowest SNR (\vs\,$\leq$ 400 \kms) shocks.
This is of course a speculation; further insight into collisionless cluster shocks may be obtained via numerical simulations
for the appropriate conditions.

\section{Summary and Future Work}
\label{sec:9}

There have been exciting advances in the study of electron-ion temperature equilibration in collisionless shocks during the 
past few years.  Perhaps most notable has been the growing realization that temperature equilibration and cosmic ray acceleration may
be intertwined processes.  Optical studies of collisionless shocks in partially neutral gas (Balmer-dominated shocks) have
shown that the electron-ion temperature equilibration is a declining function of shock speed, well characterized as
\tetp\,$\propto\,V_s^{-2}$.  This relationship most likely arises due to electron heating ahead of the shock that is 
nearly independent of shock speed above 400 \kms.  Cosmic ray precursors, with moderately amplified preshock magnetic field
and density, are the most logical sites for electron heating in SNR collisionless shocks.  The transition to fully equilibrated
SNR shocks at speeds below 400 \kms\, may be due to a less turbulent and more laminar structure at low shock speeds and Mach
numbers.  This allows the electrons to experience a more uniform cross-shock potential, and hence higher energization compared
to higher shock speeds and Mach numbers, where the shock jump is more turbulent and disordered.  The magnetosonic Mach numbers of SNR shocks
may match those in solar wind shocks if there is a approximately an order of magnitude increase in the Alfv\'en speed of
the preshock gas in SNRs compared to the average ISM value.  This is possible if there is a moderately amplified preshock 
magnetic field ($\sim$10), and is readily
provided by compression and heating in a cosmic ray precursor.  In solar wind shocks, the precursor is due to suprathermal,
non-relativistic ions, resulting in a \tetp\,$\propto\,1/V_s$ seen in the solar wind.

While Balmer-dominated shocks have allowed us to elucidate some of the physics of electron-ion temperature equilibration,
ion-neutral damping limits most of those observed to cases where the shock structure has not become strongly modified
by cosmic ray acceleration. Given this limitation, {\it electron-ion equilibration studies of fast, collisionless shocks in fully pre-ionized
gas would be highly desirable}.  Such a sample would allow the equilibration to be studied over a range of speeds
where shocks are increasingly affected by feedback from the accelerated cosmic rays. In such circumstances it is
unclear what would happen to the \tetp\, versus \vs\ relation.  If, as predicted by Amato \& Blasi (2009), Bell's non-resonant cosmic ray 
instability takes over from the resonant instability at the highest shock speeds, then additional electron heating
may occur in the fastest shocks (\vs\,$\gtrsim$5000 \kms), resulting in substantial deviation from the \tetp\,$\propto\,V_s^{-2}$ relation.
Such deviations may already have been seen in the fastest (\vs\,$\gtrsim$2000 \kms) shocks, where there is evidence that
\tetp\, does not settle down to \memp, but rather $\sim$0.03.  Other deviations may have been detected in SNR 0509$-$67.5, where
\tetp\, for a 5000 \kms\, shock has been estimated to be $\sim$0.2, substantially higher than predicted by the inverse squared relation.
However, the study of such shocks would be challenging.
Without an H$\alpha$ broad component to constrain the range of \vs, shock speeds would have to be determined by other
means, such as proper motion studies.  That would require X-ray and or UV imagery of SNRs with well-constrained distances (such as those
in the LMC or SMC), over multiple epochs.  It would also require spectroscopy of the forward shocks
in these SNRs, in order to constrain both the electron temperature (via X-ray continuum fits and and UV emission line ratios)
and the ion temperature (e.g., via He~II, C~IV, N~V and O~VI resonance lines).

An important test of our ideas concerning electron heating by cosmic ray generated waves
in a shock precursor would be to measure electron temperatures at SNR shocks exhibiting
strong cosmic ray modification and substantial magnetic field amplification ($\Delta\,B/B\,\gtrsim$100). 
Several SNRs show X-ray filaments produced by synchrotron radiation from cosmic ray electrons, 
and are generally distinct from those shocks with strong Balmer emission.
The absence of neutral material ahead of these shock means that optical and UV emission is weak,
and electron temperatures will have to be measured from X-ray spectra. Difficulties arise in
distinguishing thermal electron bremsstrahlung from cosmic ray electron synchrotron emission,
requiring data of high signal to noise. Further complications in some SNRS (e.g. Cas A) stem
from scattering of bright X-ray emission from the ejecta such that it coincides spatially with
emission from the forward shock. Such scattering may either be local, due to SNR dust, or
instrumental, due to telescope imperfections. This leaves SN 1006 as the best likely target for such
an observation, since due to its evolutionary state, only the outer layers of ejecta have 
encountered the reverse shock.

The development of missions like Solar Orbiter and Solar Probe Plus will allow {\it in situ}
measurements of shocks in the solar wind, most likely associated with coronal mass ejections, much 
closer to the Sun. These will probe a different parameter regime, where the magnetic field
pressure dominates over the gas pressure (low $\beta$, similar to ISM shocks). As such, measurements here might yield
insights into the properties of similar plasma in the precursors of SNR shocks where the magnetic
field has been amplified by cosmic rays.

\begin{acknowledgements}

P. G. acknowledges support by HST grant HST-GO-11184.07-A to Towson University.  JML acknowledges support by 
grant NNH10A009I from the NASA Astrophysics Data Analysis Program,
and by basic research funds of the Office of Naval Research.
\end{acknowledgements}

\bibliographystyle{plainnat}

\end{document}